\documentclass[11pt]{article}
\usepackage[T1]{fontenc}
\usepackage[margin=1in]{geometry}
\usepackage{amsmath,amssymb,amsthm}
\usepackage{bm}
\usepackage{algorithm,algorithmic}
\usepackage{float}
\usepackage{booktabs,tabularx,array}
\usepackage{graphicx}
\usepackage{placeins}
\usepackage{natbib}
\usepackage{tikz}
\usetikzlibrary{positioning,shapes.geometric,arrows.meta}
\usepackage[hidelinks]{hyperref}
\hypersetup{
  pdftitle={Norm-equation methods for structured group-penalized regression},
  pdfauthor={Jiayi Yu and Baisuo Jin}
}

\graphicspath{{./art/}}

\DeclareMathOperator*{\argmin}{arg\,min}

\newtheorem{theorem}{Theorem}
\newtheorem{lemma}{Lemma}
\newtheorem{corollary}{Corollary}
\newtheorem{proposition}{Proposition}

\theoremstyle{definition}
\newtheorem{assumption}{Assumption}

\newenvironment{keywords}{\par\smallskip\noindent\textbf{Keywords: }\ignorespaces}{\par\medskip}

\let\arxivOriginalLabel\label
\def\arxivCurrentPrefix{main-}
\renewcommand{\label}[1]{%
  \arxivOriginalLabel{#1}%
  \expandafter\arxivOriginalLabel\expandafter{\arxivCurrentPrefix#1}%
}

\title{Norm-equation methods for structured group-penalized regression}
\author{%
JIAYI YU\\
\small Department of Statistics and Finance, School of Management,\\
\small University of Science and Technology of China, Hefei, Anhui 230026, China\\
\small \texttt{jyj1118@mail.ustc.edu.cn}
\and
BAISUO JIN\\
\small Department of Statistics and Finance, School of Management,\\
\small University of Science and Technology of China, Hefei, Anhui 230026, China\\
\small \texttt{jbs@ustc.edu.cn}
}
\date{}

\begin{document}
\maketitle
\begin{abstract}
Group-penalized regression with correlated predictors requires block updates
for which familiar group soft thresholding formulas are generally inapplicable.
Although proximal-gradient methods can solve such updates directly, their
computational cost depends on the penalty and the design. We develop a norm-equation framework for a Euclidean group penalty combined with a finite,
closed, convex, positively homogeneous within-group penalty. Each nonzero
block minimizer is characterized by a unique scalar root, evaluated through
strongly convex inner problems without enumerating sparsity or fusion
patterns. Valid inner subgradients provide computable bounds on the scalar function and the block Karush--Kuhn--Tucker (KKT) error. These bounds allow scalar-search decisions and acceptance of approximate updates without
solving every inner problem to high accuracy. Accounting for gradient changes
during a cyclic sweep then gives a computable global KKT bound. Combined with
standard stochastic-score and restricted-curvature conditions, this bound leads to a regularization-scaled stopping rule under which approximate
estimators retain the usual estimation and prediction rates. Experiments with sparse-group and fused sparse-group regression show that computational performance depends strongly on the inner solver, with coordinate-descent and dual implementations offering advantages over direct proximal methods in some highly correlated settings.
\end{abstract}


\section{Introduction}
\label{sec:introduction}

Group-penalized regression selects prespecified sets of coefficients while
allowing structure within each selected group. The group lasso uses a
Euclidean norm \citep{yuan2006model}, the sparse-group lasso additionally
encourages coordinatewise sparsity \citep{simon2013sparse}, and fusion
penalties encourage equal coefficients along an ordering or graph
\citep{tibshirani2005fused,beer2019incorporating,stokell2021modelling,
behdin2026modelling}. We study computation for these convex estimators
when predictors within a group are correlated, retaining the penalty on
the original coefficients.

Block coordinate descent (BCD) reduces the regression problem to groupwise penalized quadratic updates
\citep{tseng2001convergence,qin2013efficient,liang2024sparsegl}. When the block Gram matrix is proportional to the identity, the exact block update can be obtained by applying the proximal operator of the within-group penalty, followed by group soft-thresholding (see Proposition~\ref{prop:isotropic}). Otherwise, these
formulas do not directly apply. General block Gram matrices can nevertheless be handled by proximal gradient (PG) and
the fast iterative shrinkage-thresholding
algorithm (FISTA) \citep{beck2009fast}; accelerated generalized gradient methods have been
used for sparse-group lasso \citep{simon2013sparse}, and smoothing proximal
gradient has been developed for structured penalties including graph-guided
fusion \citep{chen2012smoothing}. These methods use quadratic upper bounds to construct iterative updates. PG and FISTA handle general block Gram matrices through iterative proximal updates based on quadratic surrogates. In contrast, we derive an exact characterization of the block minimizer through a scalar equation and strongly convex inner problems that retain the original Gram matrix.

For the group lasso under a general design,
\citet{foygel2010exact} describe the single line search
(SLS) algorithm for exact block updates; the same
group-lasso result was obtained earlier by
\citet{puig2009multidimensional}.
They further extend this approach to the sparse-group
lasso through the signed single line search (SSLS)
algorithm, which searches over possible coefficient
sign patterns. Our characterization extends this
scalar-search perspective to a broader class of convex
within-group penalties, including fusion penalties,
without requiring enumeration of sign, support, or
fusion patterns.

The key idea is to characterize the exact block solution through a scalar equation that relates an auxiliary parameter to the Euclidean norm of the block solution. This characterization retains the original block Gram matrix without requiring a quadratic approximation. For a nonzero block, our norm-equation characterization reformulates the original block optimization problem into two linked components: a strongly convex inner optimization problem and an outer scalar equation. When the scalar satisfies the norm equation, the corresponding inner solution is exactly the minimizer of the original block problem. We establish the endpoint limits, monotonicity and uniqueness of the active
root of the norm equation for the convex, positively homogeneous penalties considered here. These properties allow the scalar equation to be solved by a bracketed search
without enumerating sign, support or fusion patterns. The monotonicity of the norm equation is closely related to classical resolvent--Yosida theory \citep{yosida1948differentiability, brezis1973operateurs}. We exploit this connection to establish the validity of
the scalar search for exact block updates.

The monotonicity of the norm equation allows a bracketed scalar search, but each evaluation requires solving a
strongly convex inner problem, whose computational cost depends on its formulation and numerical solver.
Importantly, the scalar search usually requires only determining which side of the root a trial value lies on,
rather than evaluating the norm equation
to high accuracy. We exploit this property by constructing
computable bounds on the scalar function from a valid subgradient at an approximate inner solution. These bounds allow us to update the scalar bracket, refine the
inner solution when necessary, or accept an approximate block update once its KKT error is sufficiently small.

The preceding analysis concerns individual block updates within block coordinate descent. However, subsequent block updates change the loss gradients of previously updated blocks. We derive an exact expression for this gradient
drift and use it to establish a computable KKT
error bound after a complete sweep.

Statistical theory for regularized estimators provides estimation and prediction error bounds under suitable conditions \citep{lounici2011oracle,negahban2012unified}.
This suggests that solving the optimization problem to arbitrarily high accuracy may be unnecessary, a connection also explored by \citet{agarwal2012fast}. Motivated by this observation, we develop an early stopping rule based on the global KKT bound and show that the resulting approximate
estimators retain the usual statistical rates.

Experiments with sparse-group and fused sparse-group penalties compare
norm-equation solvers with direct PG and FISTA at controlled accuracy. Suitable inner solvers can be faster for strongly correlated designs. Experiments with sparse-group and fused sparse-group regression show that computational performance depends strongly on the inner solver, with coordinate-descent and dual implementations offering advantages over direct proximal methods in some highly correlated settings.A separate experiment uses the known simulation noise to check the score condition and examines the global KKT bound, early termination and estimation accuracy as sample size increases.

Throughout the paper, $\|\cdot\|_1$ and
$\|\cdot\|_2$ denote the $\ell_1$ and Euclidean
norms, respectively. For a matrix $\bm A$,
$\bm A^\top$ and $\operatorname{range}(\bm A)$
denote its transpose and column space.
For a symmetric matrix $\bm M$,
$\lambda_{\min}(\bm M)$ denotes its smallest
eigenvalue. We use $\bm I_m$ for the
$m\times m$ identity matrix and $\partial R$
for the subdifferential of a convex function
$R$. Finally,
$\operatorname{dist}(\bm z,C)
=\inf_{\bm u\in C}\|\bm z-\bm u\|_2$
denotes the Euclidean distance from
$\bm z$ to a set $C$.

\section{Exact block updates and the norm equation}
\label{sec:norm_equation}

\subsection{The block problem and a closed-form special case}
\label{subsec:isotropic}

We consider the linear regression model
\begin{equation}
\label{eq:linear_model}
\bm y=\bm X\bm\beta^*+\bm\varepsilon,
\end{equation}
where
$\bm y\in\mathbb R^n$ is the response,
$\bm X\in\mathbb R^{n\times p}$ is the design matrix,
$\bm\beta^*\in\mathbb R^p$ is the unknown coefficient vector, and
$\bm\varepsilon\in\mathbb R^n$ is the noise vector.
The predictors are partitioned into $G$ prespecified groups,
$$
\bm X
=
(\bm X_1,\ldots,\bm X_G),
\qquad
\bm X_j\in\mathbb R^{n\times m_j},
\qquad
\sum_{j=1}^Gm_j=p,
$$
and the coefficient vector is partitioned conformably as $\bm\beta=(\bm\beta_1^\top,\ldots,\bm\beta_G^\top)^\top$.

We study the block-separable group-penalized estimator
\begin{equation}
\label{eq:general_group_problem}
\widehat{\bm\beta}
\in
\arg\min_{\bm\beta}
\left\{
\frac{1}{2n}
\|\bm y-\bm X\bm\beta\|_2^2
+
\sum_{j=1}^G
\left[
R_j(\bm\beta_j)
+
\tau_j\|\bm\beta_j\|_2
\right]
\right\},
\end{equation}
where $\tau_j>0$ and
$R_j:\mathbb R^{m_j}\to [0,\infty)$ is a convex
within-block penalty. The Euclidean norm controls selection of the entire
block, whereas $R_j$ can impose additional structure within an active block.

Several familiar estimators are contained in
\eqref{eq:general_group_problem}. For the ordinary group lasso,
$R_j=0$ and $\tau_j=\lambda\sqrt{m_j}$
\citep{yuan2006model}. For the sparse-group lasso,
$R_j(\bm b)=\lambda_1\|\bm b\|_1$ and
$\tau_j=\lambda_2\sqrt{m_j}$, so that group selection and
coordinatewise sparsity are imposed simultaneously
\citep{simon2013sparse}. A fused sparse-group penalty is obtained by
taking
$R_j(\bm b)=\lambda_1\|\bm b\|_1+\lambda_f\|\bm D_j\bm b\|_1$,
where $\bm D_j$ is a within-group difference matrix encoding the
fusion structure \citep{beer2019incorporating}. For an ordered block
$\bm b=(b_1,\ldots,b_{m_j})^\top$, a common choice is the first-difference
matrix
$$
\bm D_j
=
\begin{pmatrix}
-1 & 1 & 0 & \cdots & 0\\
0 & -1 & 1 & \ddots & \vdots\\
\vdots & \ddots & \ddots & \ddots & 0\\
0 & \cdots & 0 & -1 & 1
\end{pmatrix}
\in\mathbb R^{(m_j-1)\times m_j},
$$so that $\|\bm D_j\bm b\|_1=\sum_{k=1}^{m_j-1}|b_{k+1}-b_k|$. More generally, $\bm D_j$ may be taken as the signed incidence matrix of
a prespecified within-group graph, in which case
$\|\bm D_j\bm b\|_1$ penalizes differences between coefficients connected
by an edge.

Block coordinate descent updates one coefficient block while holding the
remaining blocks fixed. For group $j$, let $\bm r_j=\bm y-\sum_{k\ne j}\bm X_k\bm\beta_k$ be the current partial residual. The exact block update is
\begin{equation*}
\hat{\bm\beta}_j=\arg\min_{\bm b\in\mathbb R^{m_j}}
\left\{
\frac{1}{2n}
\|\bm r_j-\bm X_j\bm b\|_2^2
+
R_j(\bm b)
+
\tau_j\|\bm b\|_2
\right\}.
\end{equation*}
Expanding the squared loss, dropping terms independent of $\bm b$, and
suppressing the group index gives the generic block problem
\begin{equation}
\label{eq:block_generic}
\min_{\bm b\in\mathbb R^m}
\Phi(\bm b)= \frac12\bm b^\top\bm M\bm b -
\bm z^\top\bm b
+
R(\bm b)
+
\tau\|\bm b\|_2,
\end{equation}
where
\begin{equation*}
\bm M=\frac1n\bm X_j^\top\bm X_j,
\qquad
\bm z=\frac1n\bm X_j^\top\bm r_j.
\end{equation*}
Since $\bm M=\bm X_j^\top\bm X_j/n$ and
$\bm z=\bm X_j^\top\bm r_j/n$, we have
$\bm z\in\operatorname{range}(\bm M)$, even when $\bm X_j$ is
rank deficient.

\begin{assumption}
\label{ass:basic}
For the generic block problem, assume that
\(\bm M\succeq0\), \(\tau>0\), and that
\(R:\mathbb R^m\to[0,\infty)\) is finite valued, closed, convex, and
positively homogeneous
\[
R(t\bm b)=tR(\bm b),
\qquad
\bm b\in\mathbb R^m,\quad t\ge0.
\]
\end{assumption}

Assumption~\ref{ass:basic} applies to each within-group penalty $R_j$ in the regression problem. The optimization results do not require symmetry, that is, $R(-\bm b)=R(\bm b)$. This additional condition is imposed through the seminorm assumption only for the statistical analysis in
Section~\ref{subsec:stopping_rule}.

Let $\widehat{\bm b}$ denote the minimizer of
\eqref{eq:block_generic}. The block update may be zero or nonzero.
The zero solution is characterized by the KKT condition
$\bm z=\bm q+\tau\bm u$ for some
$\bm q\in\partial R(\bm0)$ and
$\|\bm u\|_2\le1$, or equivalently,
$\operatorname{dist}\{\bm z,\partial R(\bm0)\}\le\tau$.
We therefore focus on the active-block case
$\widehat{\bm b}\ne\bm0$. The Euclidean group norm admits
the variational identity
\begin{equation*}
\tau\|\bm b\|_2
=
\inf_{a>0}
\left\{
\frac a2\|\bm b\|_2^2
+\frac{\tau^2}{2a}
\right\}.
\end{equation*}
This identity leads to an exact two-level characterization
of a nonzero block update.

For a fixed scalar $a>0$, we define the inner problem as
the strongly convex optimization problem
\begin{equation}
\label{eq:inner_problem}
\bm b(a)
=
\arg\min_{\bm u\in\mathbb R^m}
\left\{
\frac12\bm u^\top(\bm M+a\bm I_m)\bm u
-\bm z^\top\bm u+R(\bm u)
\right\}.
\end{equation}
The outer problem seeks a scalar $a>0$
satisfying the norm equation
\begin{equation}
\label{eq:outer_problem}
\psi(a)=a\|\bm b(a)\|_2=\tau.
\end{equation}
If $a^*>0$ solves the outer problem, the corresponding
block update is $\bm b(a^*)$. The following lemma
establishes the exactness of this decomposition.

\begin{lemma}
\label{lem:norm_reduction}
Under Assumption~\ref{ass:basic} and
$\bm z\in\operatorname{range}(\bm M)$,
\eqref{eq:block_generic} has a unique minimizer
$\widehat{\bm b}$, and $\bm b(a)$ is uniquely
defined for every $a>0$.
If $\widehat{\bm b}\ne\bm0$, then
$\widehat{\bm b}=\bm b(a)$ if and only if
$\psi(a)=\tau$.
\end{lemma}
To see the equivalence, the first-order condition for
an active block is
\[
\bm0\in
\bm M\widehat{\bm b}-\bm z
+\partial R(\widehat{\bm b})
+\tau\frac{\widehat{\bm b}}
{\|\widehat{\bm b}\|_2}.
\]
Setting $a=\tau/\|\widehat{\bm b}\|_2$ gives
\begin{equation*}
\bm0\in
(\bm M+a\bm I_m)\widehat{\bm b}
-\bm z+\partial R(\widehat{\bm b}),
\end{equation*}
which is precisely the optimality condition of
\eqref{eq:inner_problem}. Conversely, if
$\psi(a)=\tau$, then $\bm b(a)\ne\bm0$
and the inner optimality condition becomes the
first-order condition of \eqref{eq:block_generic}.
For the remaining implication, if $\bm b(a)=\widehat{\bm b}\ne\bm0$,
take inner products of the two optimality conditions with $\widehat{\bm b}$.
Positive homogeneity gives $\bm q^\top\widehat{\bm b}=R(\widehat{\bm b})$
for every $\bm q\in\partial R(\widehat{\bm b})$, so subtraction yields
$a\|\widehat{\bm b}\|_2=\tau$.
The existence and uniqueness results are proved in Supplementary Section~\ref{supp-supsec:lemma1}.

The inner solution can be expressed explicitly when the within-group Gram matrix has a particular structure. We first consider $\bm M=c\bm I_m$ for $c>0$, which occurs when the predictors within a group are orthogonal and have equal squared lengths. In this case, the solution has a simple form.

Define
\begin{equation*}
\operatorname{prox}_{R}(\bm z)
=\arg\min_{\bm v\in\mathbb R^m}
\left\{
\frac12\|\bm v-\bm z\|_2^2
+
R(\bm v)
\right\},
\end{equation*}
and, with \((x)_+=\max\{x,0\}\),
\begin{equation*}
\mathbf T_\tau(\bm v)
=\left(
1-\frac{\tau}{\|\bm v\|_2}
\right)_+
\bm v,
\qquad
\mathbf T_\tau(\bm0)=\bm0.
\end{equation*}

\begin{proposition}
\label{prop:isotropic}
Under Assumption~\ref{ass:basic}, for a given $c>0$, the following statements are equivalent.

\begin{enumerate}
\item[(i)]
The within-group Gram matrix satisfies $\bm M=c\bm I_m$. 
\item[(ii)]
For every $\bm z\in\mathbb R^m$ and every $a>0$,
\begin{equation}
\label{eq:isotropic_path}
\bm b(a)=
\frac{1}{c+a}
\operatorname{prox}_{R}(\bm z).
\end{equation}
In particular, $(c+a)\bm b(a)$ remains constant as $a$ varies.

\item[(iii)]
For every $\bm z\in\mathbb R^m$ and every $\tau>0$, the exact minimizer
of \eqref{eq:block_generic} is
\begin{equation}
\label{eq:nested_threshold_general}
\widehat{\bm b}
=\frac1c\mathbf T_\tau
\left\{
\operatorname{prox}_{R}(\bm z)
\right\}.
\end{equation}
\end{enumerate}
\end{proposition}
When $\bm M=c\bm I_m$, positive homogeneity of $R$ gives
\eqref{eq:isotropic_path}. Thus changing $a$ only multiplies the
inner solution by a positive scaling factor; it does not change
its direction when it is nonzero. The converse implication is proved in Supplementary Section~\ref{supp-supsec:prop1}.

Proposition~\ref{prop:isotropic} recovers the familiar
thresholding rules as special cases. For the ordinary
group lasso, $R=0$ and
$\operatorname{prox}_R(\bm z)=\bm z$, giving the familiar group soft-thresholding operator
$$
\widehat{\bm b}=\frac1c
\left(
1-\frac{\tau}{\|\bm z\|_2}
\right)_+
\bm z.
$$
For the sparse-group lasso,
$R(\bm b)=\lambda_1\|\bm b\|_1$, and hence $\operatorname{prox}_{R}(\bm z)=S_{\lambda_1}(\bm z)$, where $S_{\lambda_1}$ denotes the coordinatewise soft-thresholding operator,
$$
S_{\lambda_1}(\bm z)
=
\bigl(
S_{\lambda_1}(z_1),\ldots,S_{\lambda_1}(z_m)
\bigr)^\top, \quad S_{\lambda_1}(z_k)
=
\operatorname{sign}(z_k)
\bigl(|z_k|-\lambda_1\bigr)_+.
$$
Therefore,
\begin{equation*}
\widehat{\bm b}
=\frac1c
\left(
1-\frac{\tau}
{\|S_{\lambda_1}(\bm z)\|_2}
\right)_+
S_{\lambda_1}(\bm z).
\end{equation*}

The covariance-regression model of
\citet{kim2026highdimensional} provides an example
where the block Gram matrix is proportional to the identity.
They consider
$\operatorname{Cov}(\bm y_i\mid\bm x_i)
=\bm B_0+\sum_{l=1}^q x_{il}\bm B_l$,
where each $\bm B_l$ is a symmetric coefficient matrix.
Writing $\bm b_l=\operatorname{vech}(\bm B_l)$,
their unconstrained estimation problem combines
elementwise sparsity and group sparsity through
$\lambda\|\bm b_l\|_1+\lambda_g\|\bm b_l\|_2$
for each nonintercept block.
Holding the remaining blocks fixed, the update
for $\bm b_l$ takes the form
\[
\min_{\bm b}
\left\{
\frac{1}{2n}
\|\bm V_l-\bm x_l\bm b^\top\|_F^2
+\lambda\|\bm b\|_1+\lambda_g\|\bm b\|_2
\right\},
\]
where $\bm x_l=(x_{1l},\ldots,x_{nl})^\top$
and $\bm V_l$ is the corresponding partial residual
matrix. Expanding the quadratic loss gives
\[
\bm M_l=c_l\bm I,\qquad
c_l=\frac{\|\bm x_l\|_2^2}{n}.
\]
Thus Proposition~\ref{prop:isotropic} applies
directly, and the nested sparse-group thresholding
rule in \eqref{eq:nested_threshold_general} gives the
exact block update. Here $\bm M_l=c_l\bm I$ follows from the statistical model
itself; no transformation of the predictors is needed.

\subsection{The norm equation for general within-group Gram matrices}
\label{subsec:norm_principle}

When $\bm M$ is not proportional to the identity, the
inner solution generally does not change by a common scaling factor. Even with $R=0$, we have $\bm b(a)=(\bm M+a\bm I_m)^{-1}\bm z$, whose direction may vary with $a$. With a nonzero within-block penalty, the nonzero coefficients,
their signs or the equalities between them may also change. However, the outer equation \eqref{eq:outer_problem} can be solved efficiently once the continuity and monotonicity of
$\psi(a)=a\|\bm b(a)\|_2$ are established.

To establish this property, we use a classical result from
resolvent--Yosida theory
\citep{yosida1948differentiability,brezis1973operateurs}. 
An operator $A$ is called monotone if, for
any two points $\bm x_1,\bm x_2$ and any
corresponding outputs
$\bm u_1\in A(\bm x_1)$ and
$\bm u_2\in A(\bm x_2)$, $\langle \bm u_1-\bm u_2,
\bm x_1-\bm x_2\rangle\ge0$. Here $A(\bm x)$ may contain more than one vector, so the condition must hold for
every choice of outputs. The subdifferential of a convex function is a monotone operator. The result below describes how the size of a regularized subgradient changes with its regularization parameter, which is exactly what is needed
for the norm equation.

\begin{lemma}
\label{lem:yosida}
Let $g:\mathbb R^m\to(-\infty,\infty]$ be proper,
closed and convex, with
$\bm0\in\operatorname{dom}(\partial g)$.
Set $A=\partial g$. For $\lambda>0$,
define the resolvent $J_{\lambda A}(\cdot)$
and Yosida approximation $A_\lambda(\cdot)$ by
\[
J_{\lambda A}(\bm x)
=(\bm I+\lambda A)^{-1}(\bm x),
\qquad
A_\lambda(\bm x)
=\frac{\bm x-J_{\lambda A}(\bm x)}{\lambda},
\]
for $\bm x\in\mathbb R^m$.
Then $\|A_\lambda(\bm0)\|_2$ is continuous
and nonincreasing in $\lambda$, and
\[
\lim_{\lambda\downarrow0}
\|A_\lambda(\bm0)\|_2
=\operatorname{dist}\{\bm0,A(\bm0)\}.
\]
\end{lemma} Here $J_{\lambda A}(\bm x)$ is the unique vector
$\bm v$ satisfying
$\bm x-\bm v\in\lambda\partial g(\bm v)$;
equivalently, it minimizes
$g(\bm v)+\|\bm v-\bm x\|_2^2/(2\lambda)$.
Thus $A_\lambda(\bm x)=(\bm x-J_{\lambda A}(\bm x))/\lambda$
is obtained from an ordinary strongly convex minimization.
We only need these quantities at $\bm x=\bm0$. The standard Yosida
properties used here are summarized in Supplementary Section~\ref{supp-supsec:yosida_lemma}.

For our block problem, take
\[
g(\bm b)=\frac12\bm b^\top\bm M\bm b
-\bm z^\top\bm b+R(\bm b),\qquad A=\partial g.
\]
The optimality condition for \eqref{eq:inner_problem} is
$-a\bm b(a)\in\partial g\{\bm b(a)\}$.
Setting $\lambda=1/a$ in the preceding definition gives
\[
\bm b(a)=J_{(1/a)A}(\bm0),
\qquad A_{1/a}(\bm0)=-a\bm b(a).
\]
Consequently,
\begin{equation}
\label{eq:yosida_identity}
\psi(a)=a\|\bm b(a)\|_2
=\|A_{1/a}(\bm0)\|_2.
\end{equation}
As $a$ increases, $1/a$ decreases. Lemma~\ref{lem:yosida}
therefore shows that $\psi(a)$ is nondecreasing.
This use of the classical Yosida result does not require the
nonzero coefficients, signs or fusion pattern of $\bm b(a)$
to remain unchanged. A direct finite-dimensional proof using only
convexity is also given in Supplementary Section~\ref{supp-supsec:direct_norm_equation}.

\begin{theorem}
\label{thm:norm_equation}
Suppose Assumption~\ref{ass:basic} holds and
$\bm z\in\operatorname{range}(\bm M)$.
Then the following statements hold.

\begin{enumerate}
\item[(i)]
The function $\psi(a)$ is continuous and
nondecreasing on $(0,\infty)$.

\item[(ii)]
Its endpoint limits are
\begin{equation}
\label{eq:psi_endpoints}
\lim_{a\downarrow0}\psi(a)=0,
\qquad
\lim_{a\to\infty}\psi(a)
=
\operatorname{dist}
\{\bm z,\partial R(\bm0)\}.
\end{equation}

\item[(iii)]
The minimizer of \eqref{eq:block_generic}
is zero if and only if
\begin{equation*}
\operatorname{dist}
\{\bm z,\partial R(\bm0)\}\le\tau.
\end{equation*}

\item[(iv)]
If
$\operatorname{dist}
\{\bm z,\partial R(\bm0)\}>\tau$,
then there exists a unique $a^*>0$ satisfying
\begin{equation*}
\psi(a^*)=\tau,
\qquad
\widehat{\bm b}=\bm b(a^*).
\end{equation*}
\end{enumerate}
\end{theorem}

The second limit in \eqref{eq:psi_endpoints} follows from
Lemma~\ref{lem:yosida}, since
$A(\bm0)=-\bm z+\partial R(\bm0)$.
The first limit follows from
$\bm z\in\operatorname{range}(\bm M)$, which ensures that
the quadratic loss is bounded below even without the additional
$a\|\bm b\|_2^2/2$ term. The zero-block condition is the KKT
condition at $\bm b=\bm0$. For a nonzero update, uniqueness of the
positive root follows from uniqueness of the block minimizer and
$a^*=\tau/\|\widehat{\bm b}\|_2$.
The Yosida-based proof of Theorem~\ref{thm:norm_equation} is given in
Supplementary Section~\ref{supp-supsec:yosida_proof}.

The continuity and monotonicity of $\psi$ allow the
outer equation to be solved by bracketing and bisection. This does not require a differentiable solution path or a fixed support, and remains valid when $\bm M$ is rank deficient. The root, together with the corresponding inner minimizer, gives the exact solution of the original
block problem. In numerical computation, the overall cost depends on the efficiency and accuracy of the inner solver.

\section{Approximate block updates and global KKT error bounds}
\label{sec:certified_computation}

\subsection{KKT error bounds for approximate block updates}
\label{subsec:certified_block}

We first establish a computable KKT error bound for an approximate block update, then extend this bound to a complete coordinate-descent sweep, and finally examine its implications for statistical estimation. The block-level analysis accounts for both the inner optimization
error and the error in the scalar norm
equation.

For an active block, Theorem~\ref{thm:norm_equation}
characterizes the exact update as
$\widehat{\bm b}=\bm b(a^*)$, where
$a^*\|\bm b(a^*)\|_2=\tau$.
In numerical implementation, neither the inner
minimization nor the scalar norm equation needs to
be solved exactly. We show that both approximations
can be controlled through computable quantities,
yielding a bound for the resulting block update.

At a trial value $a>0$, let $\widetilde{\bm b}_a$
be an approximate inner solution and choose
$\widetilde{\bm q}_a\in\partial R(\widetilde{\bm b}_a)$.
Define the inner optimality residual
\begin{equation}
\label{eq:inner_residual}
\bm r_a
=
(\bm M+a\bm I_m)\widetilde{\bm b}_a
-\bm z+\widetilde{\bm q}_a,
\qquad
\delta_a=\|\bm r_a\|_2.
\end{equation}
The departure from the scalar norm equation is
measured by
\[
\nu_a
=
\big|a\|\widetilde{\bm b}_a\|_2-\tau\big|.
\]
These quantities describe distinct aspects of the
computation: the residual \eqref{eq:inner_residual}
measures stationarity of
the inner problem at the trial value $a$, whereas
$\nu_a$ measures how closely the current pair
$(a,\widetilde{\bm b}_a)$ satisfies the norm equation.

Their combined role follows directly from the
first-order condition of the original block problem.
For $\widetilde{\bm b}_a\ne\bm0$,
\begin{equation}
\label{eq:block_residual_decomposition}
\begin{aligned}
\bm M\widetilde{\bm b}_a-\bm z+\widetilde{\bm q}_a
+\tau\frac{\widetilde{\bm b}_a}
{\|\widetilde{\bm b}_a\|_2}
\quad=
\underbrace{\bm r_a}_{\text{inner residual}}
+
\underbrace{
\bigl(\tau-a\|\widetilde{\bm b}_a\|_2\bigr)
\frac{\widetilde{\bm b}_a}
{\|\widetilde{\bm b}_a\|_2}
}_{\text{norm-equation mismatch}}.
\end{aligned}
\end{equation}
The left-hand side of \eqref{eq:block_residual_decomposition}
is a subgradient of the original block objective.
Consequently, its norm is bounded
by $\delta_a+\nu_a$, providing a computable upper bound
on the KKT error of the block update.

To guide the outer search, we must also determine
whether $\psi(a)=a\|\bm b(a)\|_2$ lies above or
below $\tau$. Since the exact inner solution
$\bm b(a)$ is generally unavailable, we construct
an interval containing $\psi(a)$.
Let
\[
\mu_a=\lambda_{\min}(\bm M)+a,
\qquad
e_a=\delta_a/\mu_a,
\]
and define
\begin{equation}
\label{eq:scalar_interval}
L_a=a\bigl(\|\widetilde{\bm b}_a\|_2-e_a\bigr)_+,
\qquad
U_a=a\bigl(\|\widetilde{\bm b}_a\|_2+e_a\bigr).
\end{equation}
Strong convexity of the inner problem ensures that
$[L_a,U_a]$ contains $\psi(a)$.
The interval in \eqref{eq:scalar_interval} therefore
provides a rigorous comparison between $\psi(a)$
and $\tau$ without requiring an
exact inner solution.

To measure violation of the block KKT conditions, define
\begin{equation*}
\mathcal R_{\rm blk}(\bm b)
=
\operatorname{dist}\bigl[
\bm0,\,
\bm M\bm b-\bm z+\partial R(\bm b)
+\tau\partial\|\bm b\|_2
\bigr].
\end{equation*}

\begin{theorem}
\label{thm:certified}
Suppose that the conditions of
Theorem~\ref{thm:norm_equation} hold and
$\operatorname{dist}\{\bm z,\partial R(\bm0)\}>\tau$.
For every $a>0$ and every inner iterate
$\widetilde{\bm b}_a$ with
$\widetilde{\bm q}_a\in\partial R(\widetilde{\bm b}_a)$,
we have
\[
\|\widetilde{\bm b}_a-\bm b(a)\|_2\le e_a,
\qquad
L_a\le\psi(a)\le U_a,
\qquad
U_a-L_a\le 2\delta_a.
\]
Moreover, if $\widetilde{\bm b}_a\ne\bm0$, then
$\mathcal R_{\rm blk}(\widetilde{\bm b}_a)
\le\delta_a+\nu_a$.
In particular, if also $L_a\le\tau\le U_a$, then
$\mathcal R_{\rm blk}(\widetilde{\bm b}_a)
\le 2\delta_a$.
\end{theorem}

At each trial value $a$, the inner approximation
provides a certified interval $[L_a,U_a]$ containing $\psi(a)$. We use this interval to determine the sign of $\psi(a)-\tau$. If $U_a<\tau$, then $a<a^*$, and we retain the part of the bracket above $a$. If $L_a>\tau$, then $a>a^*$, and we retain the part below $a$. If $\tau\in[L_a,U_a]$, the interval does not determine the sign, and we refine the inner solve
at the same trial unless the current nonzero iterate already meets the block accuracy requirement
$\delta_a+\nu_a\leq\varepsilon_{\rm blk}$.
In this case, $2\delta_a\leq\varepsilon_{\rm blk}$
provides a sufficient acceptance condition.
The same acceptance criterion is checked during
the initial bracket expansion, so that a trial can
be accepted as soon as it provides a sufficiently
accurate block solution, even if its interval does
not determine the sign of $\psi(a)-\tau$.

In particular, let $\widetilde{\bm b}_a^{(k)}$ denote
the approximate inner solution after
$k$ iterations. Even if $\|\widetilde{\bm b}_a^{(k)}
-\bm b(a)\|_2\to0$ as $k\to\infty$, the distance from zero to the subdifferential of the inner objective at $\widetilde{\bm b}_a^{(k)}$ need not converge to zero.
A counterexample is provided in Supplementary Section~\ref{supp-supsec:residual_counterexample}. Our error bounds therefore require a valid subgradient
$\widetilde{\bm q}_a\in \partial R(\widetilde{\bm b}_a)$
at the returned point, together with an explicit evaluation of the residual $\delta_a$. As $\delta_a\to0$, the interval $[L_a,U_a]$ shrinks, allowing increasingly accurate
scalar-search decisions and block updates.

The proof of the certified block bound is given in
Supplementary Section~\ref{supp-supsec:proof_certified}.
Supplementary Section~\ref{supp-supsec:pg_subgradient}
shows that an exact proximal-gradient step provides a valid subgradient at the returned inner iterate, yielding a valid residual for the bound. The interval bounds are exact-arithmetic results; their numerical evaluation is subject to floating-point rounding errors.

\subsection{Global KKT error bounds after a full sweep}
\label{subsec:global_kkt}

A block residual is evaluated at the intermediate vector where that block
was updated. Later updates change its loss gradient, even though its
coefficient vector and penalty subgradient remain fixed. We retain the
local subgradient and compute this gradient change explicitly. This yields
one valid subgradient of the full objective at the completed sweep.

Write the objective in \eqref{eq:general_group_problem} as
\begin{equation*}
F_n(\bm\beta)
=
f_n(\bm\beta)+\sum_{j=1}^G\varphi_j(\bm\beta_j),
\qquad
f_n(\bm\beta)
=
\frac{\|\bm y-\bm X\bm\beta\|_2^2}{2n},
\end{equation*}
where
$\varphi_j(\bm b)=R_j(\bm b)+\tau_j\|\bm b\|_2$
is the original block penalty.
Let $\bm\beta^{(t)}$ denote the iterate at the
beginning of sweep $t$. After updating the first
$j$ blocks, the intermediate iterate is
\[
\bm\beta^{(t,j)}
=
\left(
\bm\beta_1^{(t+1)\top},\ldots,
\bm\beta_j^{(t+1)\top},
\bm\beta_{j+1}^{(t)\top},\ldots,
\bm\beta_G^{(t)\top}
\right)^\top,
\]
with
$\bm\beta^{(t,0)}=\bm\beta^{(t)}$
and
$\bm\beta^{(t,G)}=\bm\beta^{(t+1)}$.

Once block $j$ has been updated, its coefficient
remains fixed for the rest of the sweep. Since the
penalty is block-separable, its penalty subgradient
also remains valid. The only change in the block
KKT condition therefore comes from the loss gradient.

Let
$\bm\Delta_k^{(t)}
=\bm\beta_k^{(t+1)}-\bm\beta_k^{(t)}$
and
$\bm M_{jk}=\bm X_j^\top\bm X_k/n$.
For squared loss, the gradient drift of block $j$
is exactly
\begin{equation}
\label{eq:sweep_drift}
\begin{aligned}
\bm s_j^{(t)}
=
\nabla_j f_n\{\bm\beta^{(t+1)}\}
-\nabla_j f_n\{\bm\beta^{(t,j)}\}=
\sum_{k=j+1}^G
\bm M_{jk}\bm\Delta_k^{(t)}.
\end{aligned}
\end{equation}
In particular, the drift \eqref{eq:sweep_drift} satisfies
$\bm s_G^{(t)}=\bm0$, since the last block has no
subsequent updates.

We now use the blockwise KKT error bound from
Theorem~\ref{thm:certified}. Immediately after
updating block $j$, there exists
$\bm p_j^{(t)}
\in\partial\varphi_j\{\bm\beta_j^{(t,j)}\}$
such that
\begin{equation}
\label{eq:sweep_local_residual}
\begin{aligned}
\bm g_{j,{\rm loc}}^{(t)}
=
\nabla_j f_n\{\bm\beta^{(t,j)}\}
+\bm p_j^{(t)},
\|\bm g_{j,{\rm loc}}^{(t)}\|_2
\le\varepsilon_j^{(t)}.
\end{aligned}
\end{equation}
Here $\varepsilon_j^{(t)}$ is precisely the
computable block KKT bound established in
Theorem~\ref{thm:certified}. For an accepted
nonzero update, the bound $\delta_a+\nu_a\leq\varepsilon_{\rm blk}$
allows us to take $\varepsilon_j^{(t)} = \delta_{a,j}^{(t)}+\nu_{a,j}^{(t)}$ or $\varepsilon_j^{(t)}=2\delta_{a,j}^{(t)}$
under interval-based acceptance.
For a zero block satisfying its KKT condition exactly,
the zero-block KKT condition gives
$-\nabla_j f_n\{\bm\beta^{(t,j)}\}
\in\partial\varphi_j(\bm0)$,
so $\varepsilon_j^{(t)}=0$. Thus the blockwise error bound $\varepsilon_j^{(t)}$
accounts for the KKT residual of each block at the
time of its update. For a nonzero block, this bound
is provided by Theorem~\ref{thm:certified}.
For a block set to zero through an exact optimality
check, the KKT residual is zero. If zero is accepted only within a numerical tolerance, the corresponding residual bound must also be included in $\varepsilon_j^{(t)}$.

At the end of the sweep, the selected penalty
subgradient $\bm p_j^{(t)}$ remains valid because
$\bm\beta_j^{(t+1)}=\bm\beta_j^{(t,j)}$.
Combining the local bound with the gradient
drift therefore gives the following result.

\begin{theorem}
\label{thm:sweep_global}
Suppose that a cyclic sweep visits every block
exactly once, with the blockwise KKT bound
\eqref{eq:sweep_local_residual} holding for each block.
Then there exists
$\bm g^{(t+1)}
\in\partial F_n\{\bm\beta^{(t+1)}\}$
such that
\begin{equation}
\label{eq:sweep_residual_decomposition}
\begin{aligned}
\bm g_j^{(t+1)}
=
\bm g_{j,{\rm loc}}^{(t)}
+\bm s_j^{(t)},\ \
\|\bm g_j^{(t+1)}\|_2
\le
\varepsilon_j^{(t)}
+\|\bm s_j^{(t)}\|_2,
\qquad j=1,\ldots,G.
\end{aligned}
\end{equation}
\end{theorem}

Theorem~\ref{thm:sweep_global} separates the global
KKT error into two sources: the local errors of
individual block updates and the gradient drift
induced by subsequent updates within the same sweep.
Improving the accuracy of the inner block solves
reduces the former but does not directly control
the latter. A small global KKT bound therefore
requires both sufficiently accurate block solutions
and sufficiently small changes in the subsequent
block updates. The drift term provides a computable
measure of the additional error that remains after
a complete cyclic sweep. Under the convergence of the cyclic BCD iterates, these successive block movements vanish, and hence the sweep-induced drift also tends to zero. Thus, the global KKT bound improves through both accurate block updates and the stabilization of successive BCD sweeps.

\subsection{Stopping based on the global KKT bound}
\label{subsec:stopping_rule}

Theorem~\ref{thm:sweep_global} provides a
computable bound on the global KKT error
after each completed sweep.
In statistical estimation, however,
sampling uncertainty makes it unnecessary
to solve the optimization problem to
arbitrarily high accuracy.
We therefore propose a stopping rule
that allows a nonzero KKT error while
preserving the usual convergence rates
for estimation and prediction.

For this analysis, we additionally assume that each $R_j$ is a
finite-valued seminorm. Given $\lambda>0$, write
\[
 w_j=\frac{\tau_j}{\lambda},\qquad
 \omega_j(\bm b)=\|\bm b\|_2+\frac{R_j(\bm b)}{\tau_j},\qquad
 \omega(\bm\beta)=\sum_{j=1}^G w_j\omega_j(\bm\beta_j).
\]
Then $\omega$ is a norm and the penalty in \eqref{eq:general_group_problem}
can be written as $\lambda\omega(\bm\beta)$. This additional
assumption is needed only for the statistical results below.

Using the blockwise KKT bounds $\varepsilon_j^{(t)}$ and the
gradient changes $\bm s_j^{(t)}$ from Theorem~\ref{thm:sweep_global},
define
\begin{equation}
\label{eq:sweep_bound}
\delta_{\rm sw}^{(t)}
=\max_{1\le j\le G}
\frac{\varepsilon_j^{(t)}+\|\bm s_j^{(t)}\|_2}{w_j}.
\end{equation}
Let $\omega^*$ denote the dual norm of $\omega$,
defined by $\omega^*(\bm z)
=\sup_{\omega(\bm v)\le1}\bm z^\top\bm v$. Theorem~\ref{thm:sweep_global}
implies that there exists
$\bm g^{(t+1)}\in\partial F_n\{\bm\beta^{(t+1)}\}$ satisfying
\begin{equation}
\label{eq:global_dual_bound}
\omega^*\{\bm g^{(t+1)}\}\le\delta_{\rm sw}^{(t)}.
\end{equation} The dual-norm calculation underlying \eqref{eq:global_dual_bound} is
given in Supplementary Section~\ref{supp-supsec:proof_sweep}.
Importantly, \eqref{eq:sweep_bound} uses only quantities obtained
during computation, not the unknown coefficients or noise.

We propose to stop at the first completed sweep for which
\begin{equation}
\label{eq:lambda_scaled_stopping}
\delta_{\rm sw}^{(t)}\le\xi\lambda,
\end{equation}
where $\xi\ge0$ is a chosen tolerance fraction. To explain its
statistical effect, let
$\widetilde{\bm\beta}=\bm\beta^{(t+1)}$,
$\bm h=\widetilde{\bm\beta}-\bm\beta^*$, and
$\eta=\omega^*(\bm X^\top\bm\varepsilon/n)$.
Under the regression model \eqref{eq:linear_model},
\eqref{eq:global_dual_bound} yields
\begin{equation}
\label{eq:approximate_kkt_basic}
\frac{\|\bm X\bm h\|_2^2}{2n}
\le
\bigl(\eta+\delta_{\rm sw}^{(t)}\bigr)\omega(\bm h)
+\lambda\{\omega(\bm\beta^*)-
\omega(\bm\beta^*+\bm h)\}.
\end{equation}
A derivation of the basic inequality and the associated cone condition
is given in Supplementary Section~\ref{supp-supsec:statistical_proof}.
For an exact minimizer, the corresponding inequality contains
only $\eta$ \citep{lounici2011oracle}. Thus approximate computation
adds $\delta_{\rm sw}^{(t)}$ to the stochastic error term.
If $\eta\le\gamma\lambda$ and \eqref{eq:lambda_scaled_stopping}
holds with $\theta=\gamma+\xi<1$, their sum is at most $\theta\lambda$.
The condition on $\eta$ is used to establish statistical accuracy;
its value is not required to implement the stopping rule.

We state the resulting bounds using standard conditions for
group-penalized regression \citep{negahban2012unified}.
Let $S=\{j:\bm\beta_j^*\ne\bm0\}$ and let $\mathcal M$ be the
subspace of coefficient vectors supported on the groups in $S$.
Write $\mathcal M^\perp$ for its orthogonal complement and
$\bm v_{\mathcal M}$ and $\bm v_{\mathcal M^\perp}$ for the
corresponding components of $\bm v$. Set
$C_\theta=(1+\theta)/(1-\theta)$.

\begin{assumption}
\label{ass:restricted_eigenvalue}
There exists $\kappa>0$ such that
\[
\frac{\|\bm X\bm v\|_2^2}{n}\ge\kappa\|\bm v\|_2^2
\quad\text{whenever}\quad
\omega(\bm v_{\mathcal M^\perp})
\le C_\theta\omega(\bm v_{\mathcal M}).
\]
\end{assumption}

Define
\[
\Psi(\mathcal M)
=\sup_{\bm v\in\mathcal M\setminus\{\bm0\}}
\frac{\omega(\bm v)}{\|\bm v\|_2},
\qquad \Psi(\{\bm0\})=0.
\]
This quantity describes the size of the penalty on the
true active groups.

\begin{theorem}
\label{thm:statistical_consistency}
Suppose $0\le\gamma<1$, $\eta\le\gamma\lambda$,
\eqref{eq:lambda_scaled_stopping} holds, and
Assumption~\ref{ass:restricted_eigenvalue} is satisfied,
where $\theta=\gamma+\xi<1$. Then
$\bm h=\widetilde{\bm\beta}-\bm\beta^*$ satisfies
$\omega(\bm h_{\mathcal M^\perp})
\le C_\theta\omega(\bm h_{\mathcal M})$ and
\begin{equation}
\label{eq:certified_stat_bounds}
\begin{aligned}
\|\bm h\|_2
&\le\frac{2(1+\theta)\lambda\Psi(\mathcal M)}{\kappa},\\
\frac{\|\bm X\bm h\|_2^2}{n}
&\le\frac{4(1+\theta)^2\lambda^2\Psi^2(\mathcal M)}{\kappa}.
\end{aligned}
\end{equation}
\end{theorem}
The proof is given in Supplementary
Section~\ref{supp-supsec:statistical_proof}.
Theorem~\ref{thm:statistical_consistency} shows that statistical accuracy can be controlled through a global KKT bound measured in the dual norm. This implication applies to other optimization methods whenever such a
bound is available. For the norm-equation approach, Theorems~\ref{thm:certified} and~\ref{thm:sweep_global} provide a computable bound from approximate block updates
and the gradient changes across a cyclic sweep.
The stopping tolerance affects the constants through
$\theta$, but not the convergence rates.
In particular, if the score and restricted
eigenvalue conditions hold with probability tending to one
and $\kappa$ stays bounded away from zero, then, for fixed
$\xi<1-\gamma$,

\begin{equation}
\label{eq:certified_asymptotic_rates}
\begin{aligned}
\|\widetilde{\bm\beta}_n-\bm\beta_n^*\|_2
&=O_p\{\lambda_n\Psi(\mathcal M_n)\},\\
\frac{\|\bm X_n(\widetilde{\bm\beta}_n-\bm\beta_n^*)\|_2^2}{n}
&=O_p\{\lambda_n^2\Psi^2(\mathcal M_n)\}.
\end{aligned}
\end{equation}
These are the usual statistical orders for the corresponding
exact penalized estimator.

For a concrete example, consider the ordinary group lasso problem. Suppose, conditionally on the
design, $\bm\varepsilon\sim N(\bm0,\sigma^2\bm I_n)$ and
\begin{equation}
\label{eq:group_design_bound}
\max_{1\le j\le G}\lambda_{\max}
 (\bm X_j^\top\bm X_j/n)\le C_X.
\end{equation}
For $\gamma\in(0,1)$, $\xi\in[0,1-\gamma)$ and
$\alpha\in(0,1)$, take
\begin{equation}
\label{eq:group_lambda_choice}
\lambda_n=
\frac{\sigma\sqrt{C_X}}{\gamma\sqrt n}
\left\{1+\sqrt{\frac{2\log(G/\alpha)}{m}}\right\}.
\end{equation}

\begin{corollary}
\label{cor:group_lasso_rates}
Consider the ordinary group lasso with $G$
groups of common size $m$, of which $s$
are active, and penalty parameters
$R_j=0$ and $\tau_j=\lambda_n\sqrt m$.
Suppose \eqref{eq:group_design_bound} holds.
Let $\lambda_n$ be given by
\eqref{eq:group_lambda_choice}, with
$\gamma\in(0,1)$ and $\alpha\in(0,1)$.
Assume that
Assumption~\ref{ass:restricted_eigenvalue}
holds with constant $\kappa>0$, and
the approximate estimator satisfies
\eqref{eq:lambda_scaled_stopping}
with $\xi\in[0,1-\gamma)$.
Then, conditionally on the design,
with probability at least $1-\alpha$,
\begin{equation}
\label{eq:group_explicit_bounds}
\begin{aligned}
\|\widetilde{\bm\beta}_n-\bm\beta_n^*\|_2
&\le
\frac{2(1+\theta)\lambda_n\sqrt{sm}}{\kappa},\\
\frac{\|\bm X_n(\widetilde{\bm\beta}_n
-\bm\beta_n^*)\|_2^2}{n}
&\le
\frac{4(1+\theta)^2\lambda_n^2sm}{\kappa},
\end{aligned}
\end{equation}
where $\theta=\gamma+\xi$.
\end{corollary}

The Gaussian score bound and the proof of this corollary are given in
Supplementary Section~\ref{supp-supsec:group_corollary}.

The usual group-lasso rates follow from
\eqref{eq:group_explicit_bounds}. In particular, the errors
converge to zero in probability if $\alpha=\alpha_n\to0$,
$s\{m+\log(G/\alpha_n)\}=o(n)$, and the remaining constants
stay bounded. These results concern the accuracy of a fit once
the stopping condition is met; they do not give the number of
sweeps needed to reach it.

\section{Numerical experiments}
\label{sec:numerical}

We compare numerical accuracy and computation
for sparse-group and fused sparse-group
regression. We consider two strategies for
norm-equation updates. The Exact strategy
solves the inner problem and norm equation
to high numerical accuracy. The Joint
strategy instead uses a trial pair
$(a,\widetilde{\bm b}_a)$ and the error
bounds in Theorem~\ref{thm:certified}
to jointly control the inner optimization
error and the norm-equation error.
At each trial, these bounds determine
whether to update the scalar bracket,
refine the inner solution, or accept
the approximate block update once
its KKT error is sufficiently small.

The first experiment compares Exact and
Joint strategies using coordinate descent
or proximal gradient as inner solvers,
against direct PG and FISTA, under varying
within-group correlations and group sizes. 
An additional comparison with the signed single line search
(SSLS) method of \citet{foygel2010exact}, including its
computational scaling with group size, is reported in
Supplementary Section~\ref{supp-supsec:ssls_comparison}. The second considers fused sparse-group
regression, using a box-constrained dual
method for the Joint strategy.
A separate experiment examines the global
KKT bound and the regularization-scaled
stopping rule. All comparisons use paired
Monte Carlo samples and common penalty
parameters within each setting.
Computation time and optimization residuals
are reported separately from estimation
and prediction errors.

\subsection{Sparse-group lasso}
\label{subsec:numerical_sgl}

We consider the sparse-group lasso in
\eqref{eq:general_group_problem}, with
$R_j(\bm b)=\lambda_1\|\bm b\|_1$ and
$\tau_j=\lambda_2\sqrt m$ for groups of
common size $m$.
We generate
$\bm y=\bm X\bm\beta^*+\bm\varepsilon$,
where $\bm\varepsilon\sim N(\bm0,\bm I_n)$.
There are $G=20$ groups, each containing
$m\in\{5,10,15,20\}$ predictors, with
$p=20m$ and $n=2p$.
Different groups are independent, while within-group correlations follow
$\operatorname{corr}(X_{jk},X_{j\ell})
=\rho^{|k-\ell|}$,
where $X_{jk}$ denotes the $k$th predictor in group $j$
and $\rho\in\{0.5,0.6,0.7,0.8,0.9\}$.
The design columns are centred and scaled
to norm $\sqrt n$, and the response is centred.
Groups 1, 5, 10 and 15 are active.
At $m=5$ their nonzero coefficient vectors
are, respectively,
$(0.8,0.8,0.8,0.8,0.8)$,
$(1,-0.8,1,-0.8,1)$,
$(1.2,0,0,-1.1,0)$ and
$(0,0.95,0.95,0,-0.9)$;
for larger $m$, the corresponding
five-coordinate pattern is repeated.
All other groups are zero.
We fix $\lambda_1=0.04$ and
$\lambda_2=0.035$ without tuning for
support recovery.

We compare six block solvers within the
same cyclic block coordinate descent (BCD)
algorithm. Four solvers use the norm-equation
framework, with two strategies for the
inner and scalar problems.
The Exact strategy solves both problems
to high numerical accuracy.
The joint strategy applies
Theorem~\ref{thm:certified} to an approximate
inner solution $\widetilde{\bm b}_a$.
Using a subgradient
$\widetilde{\bm q}_a\in
\partial R(\widetilde{\bm b}_a)$,
it computes the inner residual $\delta_a$
and the bounds $[L_a,U_a]$ on $\psi(a)$.
If $U_a<\tau$ or $L_a>\tau$, the scalar
bracket can be updated without an accurate
inner solution. Otherwise, the inner
solution is refined unless the current
block already satisfies the prescribed
KKT tolerance. Thus, the Joint strategy
avoids unnecessary high-accuracy inner
solves while controlling the error of
the returned block update.

For each strategy, we use coordinate descent
(CD) or proximal gradient (PG) to solve
the strongly convex inner lasso problem,
giving NormEq--Exact--CD,
NormEq--Joint--CD, NormEq--Exact--PG
and NormEq--Joint--PG.
The remaining two methods, Direct PG and
Direct FISTA, solve the original block
problem using proximal gradient and
accelerated proximal gradient, respectively.
Both use the sparse-group proximal operator.
Direct FISTA employs an adaptive restart,
and warm starts are used across inner
iterations, scalar trials and BCD sweeps
where applicable.

All six implementations start from zero
and are required to attain
\begin{equation}
\label{eq:numerical_kkt_target}
\mathcal R_{\rm KKT}(\widehat{\bm\beta})
:=\max_j\operatorname{dist}\bigl\{\bm0,
\nabla_j f_n(\widehat{\bm\beta})+
\partial\varphi_j(\widehat{\bm\beta}_j)\bigr\}
\le 2\times10^{-5},
\end{equation}
where
$f_n(\bm\beta)=
\|\bm y-\bm X\bm\beta\|_2^2/(2n)$
and
$\varphi_j(\bm b)=
R_j(\bm b)+\tau_j\|\bm b\|_2$.
Each $(m,\rho)$ setting has 100 paired
replications. Running time includes
block optimization and convergence checks,
but excludes common design generation
and Gram-matrix preprocessing.

\begin{table}[htbp]
\centering
\caption{Computation for sparse-group lasso under varying
group dimensions and within-group correlations.
Entries are mean (standard deviation) running times
in milliseconds over 100 paired replications.
All methods attain the common KKT target in
\eqref{eq:numerical_kkt_target}.}
\label{tab:sgl-comparison}

\scriptsize
\setlength{\tabcolsep}{2pt}
\renewcommand{\arraystretch}{1.12}

\begin{tabularx}{\textwidth}{
@{}cc*{6}{>{\centering\arraybackslash}X}@{}
}
\toprule
& & \multicolumn{4}{c}{NormEq}
    & \multicolumn{2}{c}{Direct} \\
\cmidrule(lr){3-6}
\cmidrule(lr){7-8}
$m$ & $\rho$
& Exact--CD & Joint--CD
& Exact--PG & Joint--PG
& PG & FISTA \\
\midrule
5 & 0.5 & 1.63 (0.36) & 1.10 (0.20) & 3.17 (0.75) & 1.94 (0.50) & 1.40 (0.26) & 1.52 (0.19) \\
5 & 0.6 & 1.67 (0.37) & 1.15 (0.37) & 3.79 (0.75) & 2.16 (0.37) & 1.54 (0.22) & 1.60 (0.20) \\
5 & 0.7 & 1.70 (0.29) & 1.15 (0.25) & 4.69 (0.79) & 2.58 (0.53) & 1.81 (0.28) & 1.72 (0.28) \\
5 & 0.8 & 1.73 (0.33) & 1.14 (0.20) & 5.71 (0.96) & 2.96 (0.53) & 2.02 (0.27) & 1.71 (0.26) \\
5 & 0.9 & 1.66 (0.35) & 1.04 (0.15) & 6.44 (0.91) & 3.15 (0.49) & 2.14 (0.27) & 1.50 (0.17) \\
\addlinespace[5pt]
10 & 0.5 & 2.27 (0.50) & 1.84 (0.44) & 3.70 (0.80) & 2.53 (0.46) & 2.21 (0.50) & 2.28 (0.38) \\
10 & 0.6 & 2.50 (0.51) & 1.93 (0.52) & 5.01 (1.38) & 3.10 (0.56) & 2.60 (0.54) & 2.53 (0.51) \\
10 & 0.7 & 2.53 (0.38) & 1.96 (0.37) & 6.38 (1.14) & 3.61 (0.45) & 2.86 (0.64) & 2.49 (0.29) \\
10 & 0.8 & 2.59 (0.37) & 1.93 (0.26) & 8.41 (1.11) & 4.52 (0.62) & 3.46 (0.46) & 2.62 (0.29) \\
10 & 0.9 & 2.58 (0.52) & 1.86 (0.40) & 11.09 (1.57) & 5.48 (0.75) & 4.01 (0.50) & 2.53 (0.49) \\
\addlinespace[5pt]
15 & 0.5 & 4.26 (0.48) & 3.87 (0.49) & 5.70 (0.68) & 4.53 (0.49) & 4.34 (0.84) & 4.48 (1.18) \\
15 & 0.6 & 4.22 (0.51) & 3.71 (0.32) & 6.62 (0.91) & 4.79 (0.39) & 4.41 (0.35) & 4.31 (0.45) \\
15 & 0.7 & 4.50 (0.45) & 3.94 (0.58) & 8.56 (0.88) & 5.73 (0.49) & 5.08 (0.49) & 4.59 (0.41) \\
15 & 0.8 & 4.85 (0.66) & 4.01 (0.52) & 12.23 (1.40) & 7.26 (0.72) & 5.98 (0.65) & 4.91 (0.61) \\
15 & 0.9 & 4.95 (0.58) & 3.91 (0.49) & 17.35 (1.83) & 9.22 (0.96) & 7.42 (0.77) & 4.89 (0.60) \\
\addlinespace[5pt]
20 & 0.5 & 6.36 (0.65) & 5.83 (0.56) & 7.68 (0.89) & 6.46 (0.69) & 6.29 (0.64) & 6.27 (0.62) \\
20 & 0.6 & 6.67 (1.22) & 6.13 (0.96) & 9.09 (1.27) & 7.37 (1.51) & 6.84 (0.83) & 6.72 (0.93) \\
20 & 0.7 & 6.46 (0.59) & 5.80 (0.57) & 10.72 (1.25) & 7.68 (0.60) & 7.10 (0.61) & 6.44 (0.58) \\
20 & 0.8 & 6.81 (0.57) & 5.91 (0.43) & 14.55 (1.04) & 9.29 (0.60) & 8.14 (0.93) & 6.70 (0.54) \\
20 & 0.9 & 7.09 (0.94) & 5.89 (0.93) & 21.67 (2.92) & 12.90 (4.48) & 10.31 (1.44) & 6.79 (1.12) \\
\bottomrule
\end{tabularx}
\end{table}

Table~\ref{tab:sgl-comparison} shows that
NormEq--Joint--CD has the smallest mean running time in all 20 settings.
Joint certification reduces computation relative to the corresponding
Exact version for both CD and PG inner solvers. However, the gain depends
on the inner solver: NormEq--Joint--PG is generally less competitive than
the CD implementation, despite using the same scalar accuracy rule.
Direct PG becomes slower as within-group correlation increases,
whereas Direct FISTA is more stable and remains competitive for larger
$m$. 

\begin{figure}[htbp]
\centering
\includegraphics[width=0.92\textwidth]{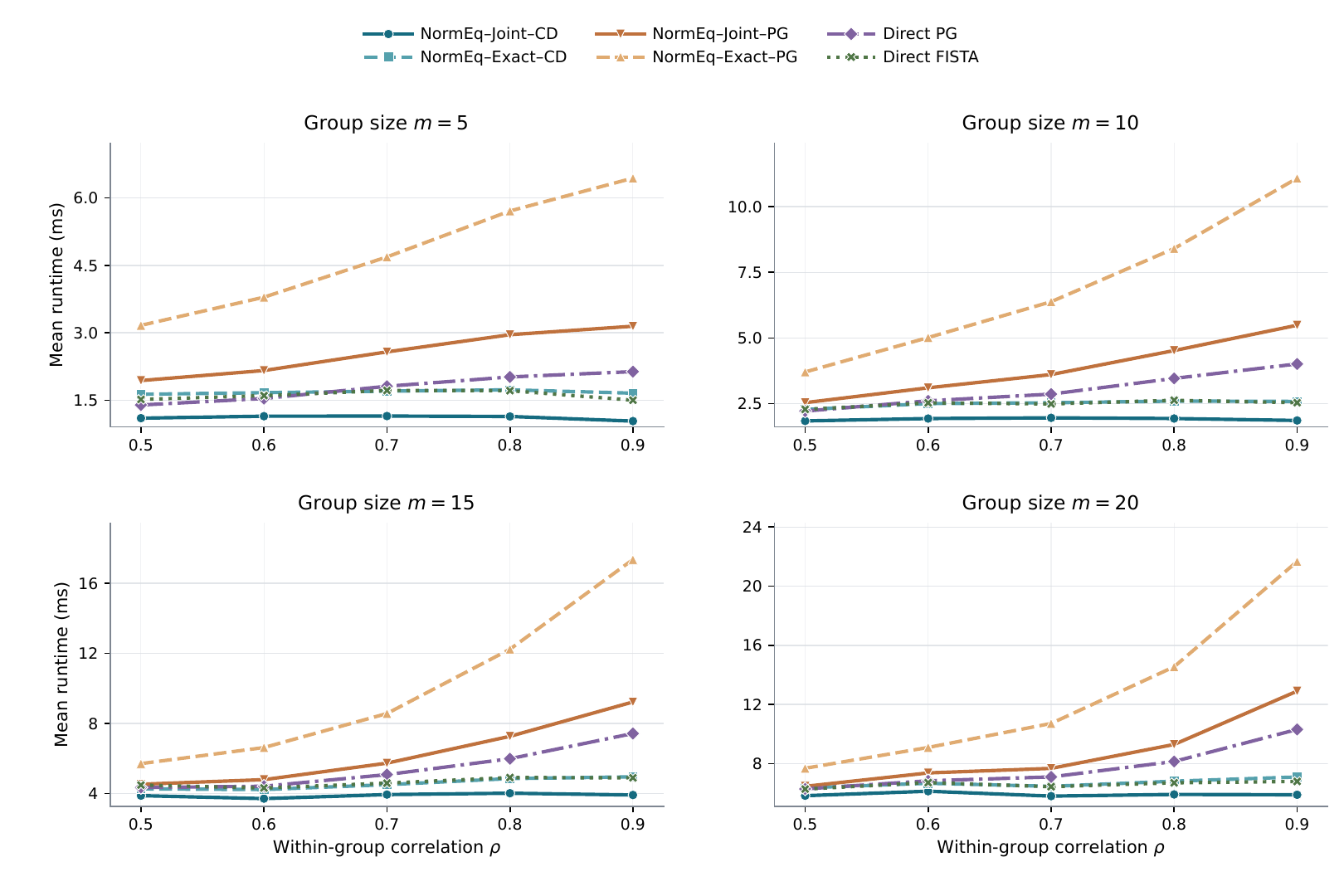}
\caption{Mean running time against within-group correlation for four group sizes. Each point averages 100 paired replications.}
\label{fig:sgl-runtime}
\end{figure}

The six methods return essentially the same estimator on each
replication: the largest difference between the maximum and minimum
objective values across the six fits is $1.14\times10^{-9}$.
All 12\,000 runs reach the common KKT target.

For example, at $\rho=0.9$, the mean (standard deviation)
coefficient $\ell_2$ errors for NormEq--Joint--CD and
Direct FISTA are identical to three decimal places:
1.860 (0.090) at $m=5$ and 3.987 (0.093) at $m=20$.
These results illustrate that the computational differences between the methods do not translate into appreciable differences in estimation accuracy.
\FloatBarrier

\subsection{Fused sparse-group regression}
\label{subsec:numerical_fsgl}

We next consider fused sparse-group regression
based on \citet{beer2019incorporating}.
There are $G=16$ groups of 25 predictors
($p=400$), each with a $5\times5$
nearest-neighbour fusion graph.
Only group 7 is active, with all 25
coefficients equal to 3; the remaining
coefficients are zero.
We retain the 640 within-group fusion
edges and omit cross-group edges,
preserving the block-separable penalty.
We generate Gaussian predictors with
independent groups and within-group
correlation $\operatorname{corr}(X_{jk},X_{j\ell})
=\rho^{|k-\ell|}$.
The predictor columns and response are
centred, and
$\bm\varepsilon\sim N(\bm0,4\bm I_n)$.
Thus $\rho=0$ corresponds to independent
predictors, while positive $\rho$
introduces within-group correlation.

We minimize the normalized objective
\begin{equation}
\label{eq:numerical_fsgl_model}
\begin{aligned}
F_n(\bm\beta)
=&\frac{\|\bm y-\bm X\bm\beta\|_2^2}{2n}
+\lambda_1\|\bm\beta\|_1
+\lambda_f\sum_{j=1}^G
\|\bm D_j\bm\beta_j\|_1\\
&+\lambda_g\sum_{j=1}^G
\sqrt{25}\|\bm\beta_j\|_2,
\end{aligned}
\end{equation}
where $\bm D_j$ is the incidence matrix
of the within-group fusion graph.
The original formulation of
\citet{beer2019incorporating} uses an
unnormalized squared-error loss.
Dividing their objective by $n=50$
gives the corresponding normalized
penalty parameters
$\lambda_1=\lambda_g=0.127978$
and $\lambda_f=0.255955$.
We keep these parameters fixed across
all settings without retuning for
correlation or sample size.

The main experiment uses
$(\rho,n)=(0.5,500),(0.6,600),
(0.7,800),(0.8,1000)$ and $(0.9,1200)$,
with $p=400$ fixed.
To separate the effect of correlation
from that of sample size, we also
compare $\rho\in\{0.5,0.7,0.9\}$
at a fixed $n=800$.
The $(\rho,n)=(0.7,800)$ setting
is shared by these comparisons.
We additionally include $(\rho,n)=(0,50)$
as a reference based on the original
independent-design setting.
Each distinct setting has 100 paired
Monte Carlo replications.

We compare three cyclic BCD implementations.
NormEq--Joint--Dual applies the approximate
block updates of
Theorem~\ref{thm:certified}, using a
box-constrained dual quadratic programme
to solve the inner problem.
The computed inner subgradient and residual
are checked to guide scalar-bracket updates
and block acceptance.
Direct PG and Direct FISTA instead solve
the original fused sparse-group block
using proximal gradient and accelerated
proximal gradient with restart.
Both methods compute their proximal steps
using the same fused proximal subproblem
followed by Euclidean group shrinkage.
Primal and dual variables are warm-started
where applicable.

We compare optimization accuracy using
the common PG mapping
\begin{equation}
\label{eq:numerical_pg_mapping}
\mathcal R_{\rm PG}(\bm\beta)
=\frac1t\max_j
\left\|
\left[
\bm\beta-\operatorname{prox}_{tP}
\{\bm\beta-t\nabla f_n(\bm\beta)\}
\right]_j
\right\|_2,
\qquad
t=\frac1{2\lambda_{\max}
(\bm X^\top\bm X/n)},
\end{equation}
where $P=F_n-f_n$.
The finite-accuracy fused proximal
calculation is accounted for in an
upper bound on this mapping.
All methods target
$\mathcal R_{\rm PG}\le10^{-4}$
with a maximum of 60 complete BCD sweeps.
NormEq--Joint--Dual additionally checks
the original nonsmooth KKT error at the
returned point.
Running times include optimization and
convergence checks, but exclude data
generation and independent post-fit
diagnostics.

We report coefficient error
$\|\widehat{\bm\beta}-\bm\beta^*\|_2$
and prediction MSE
$\|\bm X_{\rm test}
(\widehat{\bm\beta}-\bm\beta^*)\|_2^2
/n_{\rm test}$
on an independent test design,
excluding additional test noise.
For structural recovery, let
$\mathcal I=\{i:\beta_i^*\ne0\}$ and
$\widehat{\mathcal I}
=\{i:|\widehat\beta_i|>10^{-3}\}$
denote the true and estimated coefficient supports.
The support F1 score is
\[
\mathrm{F1}
=\frac{2|\mathcal I\cap\widehat{\mathcal I}|}
{|\mathcal I|+|\widehat{\mathcal I}|}.
\]
Exact group recovery occurs when
\[
\{j:\|\widehat{\bm\beta}_j\|_2>10^{-3}\}
=
\{j:\|\bm\beta_j^*\|_2>0\}.
\]
The recovery percentage is the proportion
of successful fits satisfying this condition.

\begin{table}[htbp]
\centering
\caption{Computation, estimation and recovery
for fused sparse-group regression
(100 paired replications per setting).
Times and errors are means (standard
deviations) over successful fits.
F1 scores are means, and group recovery
is reported as a percentage of successful
fits. NormEq denotes NormEq--Joint--Dual;
FISTA and PG are the direct methods.}
\label{tab:fsgl-main}
\scriptsize
\setlength{\tabcolsep}{3pt}
\renewcommand{\arraystretch}{1.15}
\resizebox{\textwidth}{!}{%
\begin{tabular}{@{}ccccccccc@{}}
\toprule
& & \multicolumn{3}{c}{Running time (s)}
& \multicolumn{4}{c}{Estimation and recovery}\\
\cmidrule(lr){3-5}
\cmidrule(lr){6-9}
$\rho$ & $n$
& NormEq & FISTA & PG
& $\ell_2$ error & Prediction MSE
& Support F1 & Group recovery\\
\midrule
0.5 & 500
& 0.467 (0.068) & 0.454 (0.074)
& 0.563 (0.086)
& 0.452 (0.047) & 0.587 (0.126)
& 1.000 & 100\% \\
0.6 & 600
& 0.461 (0.064) & 0.469 (0.057)
& 0.649 (0.096)
& 0.349 (0.049) & 0.458 (0.130)
& 1.000 & 100\% \\
0.7 & 800
& 0.474 (0.066) & 0.523 (0.070)
& 0.825 (0.114)
& 0.252 (0.034) & 0.327 (0.086)
& 1.000 & 100\% \\
0.8 & 1000
& 0.490 (0.062) & 0.599 (0.081)
& 1.170 (0.170)
& 0.176 (0.025) & 0.233 (0.066)
& 1.000 & 100\% \\
0.9 & 1200
& 0.523 (0.071) & 0.775 (0.135)
& 1.958 (0.363)
& 0.103 (0.019) & 0.137 (0.050)
& 1.000 & 100\% \\
\bottomrule
\end{tabular}}
\end{table}

The estimation and recovery results in
Table~\ref{tab:fsgl-main} are based on
successful NormEq--Joint--Dual fits.
The direct methods give nearly identical
coefficient and prediction errors.
All successful fits attain the common
PG target, and both direct methods
attain it in all 100 replications
of every setting.

At $(\rho,n)=(0.7,800)$, one
NormEq--Joint--Dual run could not
complete the required inner residual
verification within the permitted
refinements, preventing a verified
scalar-search decision.
This run is excluded from the reported
time and estimation summaries.
The group recovery rate of 100\%
at this setting therefore refers to
the 99 successful fits.
All 100 NormEq runs succeed in the
other settings.

\begin{table}[htbp]
\centering
\caption{Running times under a fixed sample
size ($n=800$), with the independent-design
reference ($n=50,\rho=0$).
Entries are means (standard deviations)
in seconds over successful fits from
100 paired replications per setting.
The speed ratio uses jointly successful
paired runs.
NormEq denotes NormEq--Joint--Dual.}
\label{tab:fsgl-controls}
\small
\setlength{\tabcolsep}{6pt}
\renewcommand{\arraystretch}{1.12}
\begin{tabular}{@{}cccccc@{}}
\toprule
$\rho$ & $n$
& NormEq (s) & Direct FISTA (s)
& Direct PG (s) & FISTA/NormEq \\
\midrule
0.5 & 800
& 0.431 (0.046)
& 0.384 (0.054)
& 0.448 (0.060) & 0.89\\
0.7 & 800
& 0.474 (0.066)
& 0.523 (0.070)
& 0.825 (0.114) & 1.10\\
0.9 & 800
& 0.657 (0.087)
& 1.070 (0.142)
& 2.959 (0.421) & 1.63\\
\midrule
0.0 & 50
& 1.785 (0.399)
& 1.650 (0.366)
& 2.088 (0.402) & 0.92\\
\bottomrule
\end{tabular}
\end{table}

Tables~\ref{tab:fsgl-main}
and~\ref{tab:fsgl-controls} show that
the relative computational performance
depends on within-group correlation.
NormEq--Joint--Dual and Direct FISTA
have comparable running times at moderate
correlation, while the former becomes
faster as correlation increases.
At $\rho=0.9$, FISTA takes 1.48 times
as long as NormEq--Joint--Dual in the
main experiment and 1.63 times as long
at fixed $n=800$.
The fixed-$n$ comparison shows that
this relative improvement is not solely
due to changes in sample size.

The successful fits give nearly identical
estimation and prediction accuracy,
with perfect group recovery in the main
settings. In the independent-design
reference, the mean support F1 is 0.996,
and exact group recovery is 96\%.
These results illustrate the computational
advantage of the norm-equation approach
under stronger correlation, while direct
proximal methods remain competitive
when correlation is lower.

\FloatBarrier

\subsection{Sweep-level KKT control and stopping}
\label{subsec:sweep_drift}

The preceding experiments compare block solvers at a common numerical
accuracy. We now examine the bound in Theorem~\ref{thm:sweep_global}
and the stopping rule \eqref{eq:lambda_scaled_stopping}. In particular,
we use the known simulation errors to evaluate the stochastic score
in Theorem~\ref{thm:statistical_consistency}, rather than inferring
its size from the observed estimation error. We separately check
whether the sweep-level bound controls the global KKT residual and
whether stopping at a regularization-scaled threshold preserves
estimation and prediction accuracy.

We consider sparse-group lasso with $G=20$ groups of size $m=10$
($p=200$) and $n\in\{400,800,1600,3200\}$. The design has
independent groups and within-group AR(1) correlation $0.5$;
predictor columns are centred and standardized, and
$\bm\varepsilon\sim N(\bm0,\bm I_n)$. Four groups are active,
with the dense and partially sparse five-coordinate coefficient
patterns used in Section~\ref{subsec:numerical_sgl}, each repeated
twice. We set
\[
\lambda_{2,n}=4/\sqrt n,\qquad
\lambda_{1,n}=0.4\lambda_{2,n},\qquad
\tau_j=\lambda_{2,n}\sqrt m.
\]
The same 100 realizations at each $n$ are used to compare stopping
thresholds. One cyclic NormEq--Joint--CD trajectory is run per
realization, with certified block updates at tolerance
$3\times10^{-6}$. We record the first completed sweep satisfying
$\delta_{\rm sw}\le\xi\lambda_{2,n}$, for
$\xi=0.10,0.05$ and $0.001$. The smallest value provides a tighter
numerical reference, not an exact optimization oracle. Prediction
MSE is evaluated on 1000 independently generated test covariates
and excludes additional test noise.

For this model, the norm in Section~\ref{subsec:stopping_rule} is
\[
\omega(\bm\beta)
=\sum_{j=1}^G\sqrt m\,\|\bm\beta_j\|_2
+\frac{\lambda_{1,n}}{\lambda_{2,n}}\|\bm\beta\|_1.
\]
Because $\bm\varepsilon$ is known in each simulation, the score
$\eta=\omega^*(\bm X^\top\bm\varepsilon/n)$ can be computed
directly. More specifically, with $S_c$ denoting coordinatewise
soft thresholding,
\begin{equation}
\label{eq:numerical_score_dual}
\eta=\max_j\inf\left\{u\ge0:
\left\|S_{u\lambda_{1,n}/\lambda_{2,n}}
\left(\frac{\bm X_j^\top\bm\varepsilon}{n}\right)\right\|_2
\le u\sqrt m\right\}.
\end{equation}
We also evaluate $\kappa_n=\lambda_{\min}(\bm X^\top\bm X/n)$,
which gives a valid, although conservative, restricted-curvature
constant. For the four active groups,
$\Psi(\mathcal M)=1.4\sqrt{4m}$.

First, we compare the bound $\delta_{\rm sw}$ against a direct
end-of-sweep KKT diagnostic,
\[
\mathcal R_2(\bm\beta)
=\max_j\frac{1}{\sqrt m}
\operatorname{dist}\!\left\{\bm0,
\nabla_j f_n(\bm\beta)+
\partial\varphi_j(\bm\beta_j)\right\}.
\]
This diagnostic is distinct from the dual-norm residual in
\eqref{eq:global_dual_bound}, but satisfies
$\mathcal R_2\{\bm\beta^{(t+1)}\}\le\delta_{\rm sw}^{(t)}$.
Table~\ref{tab:sweep_bound} displays the first three sweeps at
$n=800$. The local contributions are small; the larger gradient
changes caused by later block updates dominate the bound. The
computed global KKT diagnostic is below the bound in every
replication, including for the other three sample sizes.

\begin{table}[htbp]
\centering
\caption{Local block errors, subsequent gradient drift and
end-of-sweep KKT control at $n=800$ (100 replications).
Entries are means (standard deviations). The Local column is
reported in units of $10^{-6}$; the other columns use the
normalized scale of \eqref{eq:sweep_bound}.}
\label{tab:sweep_bound}
\small
\setlength{\tabcolsep}{4pt}
\renewcommand{\arraystretch}{1.08}
\begin{tabularx}{\textwidth}{@{}c*{4}{>{\centering\arraybackslash}X}@{}}
\toprule
Sweep & Local ($10^{-6}$) & Drift & $\delta_{\rm sw}$
& $\mathcal R_2$ \\
\midrule
1 & 0.822 (0.108) & 0.11964 (0.01958) & 0.11964 (0.01958) & 0.10480 (0.02643) \\
2 & 0.817 (0.104) & 0.00893 (0.00194) & 0.00893 (0.00194) & 0.00725 (0.00217) \\
3 & 0.805 (0.114) & 0.00057 (0.00022) & 0.00057 (0.00022) & 0.00051 (0.00024) \\
\bottomrule
\end{tabularx}
\end{table}

We next check the conditions underlying the statistical bound.
Across all 400 data sets, the realized score satisfies
$\eta/\lambda_{2,n}\le0.45$ (the maximum is $0.445$), while
$\kappa_n>0$ (the smallest observed value is $0.0434$).
Thus the score and curvature assumptions are satisfied in
these realizations, and each stopping point with $\xi\le0.10$
satisfies $(\eta+\delta_{\rm sw})/\lambda_{2,n}<1$.
We evaluate both sides of \eqref{eq:approximate_kkt_basic}
and the estimation and prediction bounds in
\eqref{eq:certified_stat_bounds}, using the realized
$\eta$, $\delta_{\rm sw}$, $\kappa_n$ and
$\Psi(\mathcal M)$. All 1200 stopping points satisfy these
inequalities numerically. The bounds obtained from the full Gram
minimum eigenvalue are conservative and are not intended as
accurate numerical predictions of the errors.

\begin{table}[htbp]
\centering
\caption{Estimation, prediction and computation under regularization-scaled
stopping (100 paired replications for each $n$).
Entries are means (standard deviations). Running time is reported
in milliseconds. The value $\xi=0.001$ gives a tighter reference
along the same NormEq--Joint--CD trajectory. Prediction MSE excludes test noise.}
\label{tab:stopping-comparison}
\scriptsize
\setlength{\tabcolsep}{2pt}
\renewcommand{\arraystretch}{1.10}
\begin{tabularx}{\textwidth}{@{}cc*{5}{>{\centering\arraybackslash}X}@{}}
\toprule
$n$ & $\xi$ & Sweeps & Time (ms) & $\delta_{\rm sw}/\lambda_{2,n}$
& $\ell_2$ error & Prediction MSE \\
\midrule
400 & 0.100 & 2.09 (0.29) & 3.908 (0.800) & 0.061 (0.022) & 2.536 (0.145) & 3.752 (0.473) \\
    & 0.001 & 4.08 (0.27) & 7.973 (1.132) & 0.0005 (0.0002) & 2.536 (0.144) & 3.746 (0.462) \\
\addlinespace[2pt]
800 & 0.100 & 2.01 (0.10) & 4.150 (0.765) & 0.062 (0.014) & 1.885 (0.089) & 1.968 (0.214) \\
    & 0.001 & 4.00 (0.00) & 8.432 (1.154) & 0.0003 (0.0001) & 1.883 (0.089) & 1.964 (0.211) \\
\addlinespace[2pt]
1600 & 0.100 & 2.02 (0.14) & 4.344 (0.621) & 0.057 (0.016) & 1.348 (0.048) & 0.980 (0.084) \\
     & 0.001 & 3.99 (0.10) & 8.907 (0.714) & 0.0001 (0.0001) & 1.347 (0.047) & 0.978 (0.082) \\
\addlinespace[2pt]
3200 & 0.100 & 2.00 (0.00) & 5.640 (0.619) & 0.049 (0.012) & 0.961 (0.031) & 0.485 (0.032) \\
     & 0.001 & 3.87 (0.34) & 11.322 (1.252) & 0.0001 (0.0002) & 0.960 (0.030) & 0.484 (0.032) \\
\bottomrule
\end{tabularx}
\end{table}

Table~\ref{tab:stopping-comparison} shows that the
regularization-scaled rule reaches its threshold after about two
sweeps, compared with about four for the tighter reference, while
the estimation and prediction errors remain close on the paired
data sets. The intermediate threshold $\xi=0.05$ gives similar
results. As $n$ increases, both errors decrease: a descriptive
log--log fit of the four mean errors at $\xi=0.10$ gives slopes
$-0.469$ for $\ell_2$ error and $-0.986$ for prediction MSE.
These patterns agree with the orders $n^{-1/2}$ and $n^{-1}$
from \eqref{eq:certified_asymptotic_rates} when the number and
sizes of groups are fixed and
$\lambda_{2,n}\asymp n^{-1/2}$. This simulation checks the
conditions and bounds on realized samples and illustrates the
stopping rule; it does not constitute an independent proof of the asymptotic assertion.

\FloatBarrier

\section{Discussion}
\label{sec:discussion}

The norm-equation approach provides an alternative to direct proximal-gradient methods for structured group-penalized regression with correlated predictors. Its principal contribution is an exact scalar characterization that accommodates different inner solvers, together with computable error bounds that connect approximate block updates to global optimality and statistical accuracy. 

The proposed approach does not guarantee
a computational advantage over direct
proximal methods in all settings.
In particular, its efficiency depends on
obtaining sufficiently accurate inner
optimality residuals, which can be
challenging for nonsmooth penalties.
Developing more efficient inner solvers
and extending the framework to more general
structured penalties are directions for
future research.

\bibliographystyle{plainnat}
\bibliography{paper-ref}

\clearpage
\setcounter{section}{0}
\setcounter{subsection}{0}
\setcounter{equation}{0}
\setcounter{table}{0}
\setcounter{figure}{0}
\setcounter{algorithm}{0}
\renewcommand{\thesection}{S\arabic{section}}
\renewcommand{\thesubsection}{\thesection.\arabic{subsection}}
\renewcommand{\theequation}{S\arabic{equation}}
\renewcommand{\thetable}{S\arabic{table}}
\renewcommand{\thefigure}{S\arabic{figure}}
\renewcommand{\thealgorithm}{S\arabic{algorithm}}
\renewcommand{\theHsection}{supp.\arabic{section}}
\renewcommand{\theHsubsection}{supp.\arabic{section}.\arabic{subsection}}
\renewcommand{\theHequation}{supp.\arabic{equation}}
\renewcommand{\theHfigure}{supp.\arabic{figure}}
\renewcommand{\theHtable}{supp.\arabic{table}}
\providecommand{\theHalgorithm}{}
\renewcommand{\theHalgorithm}{supp.\arabic{algorithm}}
\def\arxivCurrentPrefix{supp-}

\begin{center}
  {\Large\bfseries Supplementary material for\par
  ``Norm-equation methods for structured group-penalized regression''\par}
  \vspace{1em}
  JIAYI YU \quad and \quad BAISUO JIN
\end{center}

\raggedbottom
This supplement is organized as follows. Section~\ref{supsec:block_geometry}
contains proofs for the exact block reduction, the isotropic characterization
and the norm-equation theorem. Section~\ref{supsec:inner_certificates} proves
the block KKT bound, gives a counterexample showing that convergence of inner
iterates need not imply a vanishing KKT residual, and constructs a valid
subgradient from an exact proximal-gradient step. Section~\ref{supsec:global_proofs}
contains the proof of the sweep-level KKT bound and the proofs of the
estimation and prediction error bounds, including the group-lasso
corollary. Section~\ref{supsec:implementation} gives the SSLS and NormEq
algorithms, the inner-solver constructions, and the cyclic BCD stopping
rules. The additional SSLS comparison and group-size scaling experiment
are reported in Section~\ref{supsec:ssls_comparison}; proximal-gradient
accuracy and evaluation measures are summarized in
Section~\ref{supsec:metrics}. The theorem and lemma numbers refer to the
main paper; supplementary sections, equations, tables, figures and
algorithms use the prefix S.

\section{Proofs for exact block geometry and the norm equation}
\label{supsec:block_geometry}

\subsection{Proof of Lemma~\ref{main-lem:norm_reduction}}
\label{supsec:lemma1}

Write $\Phi$ for the generic block objective in the main paper and
$\bm M^+$ for the Moore--Penrose inverse. Since
$\bm z\in\operatorname{range}(\bm M)$,
\[
\tfrac12\bm b^\top\bm M\bm b-\bm z^\top\bm b
=\tfrac12(\bm b-\bm M^+\bm z)^\top\bm M
 (\bm b-\bm M^+\bm z)-\tfrac12\bm z^\top\bm M^+\bm z.
\]
Consequently $\Phi(\bm b)\ge
\tau\|\bm b\|_2-\bm z^\top\bm M^+\bm z/2$.
The objective is continuous and coercive, so it attains its minimum,
including when $\bm M$ is singular.

To prove uniqueness, suppose that $\bm b_1$ and $\bm b_2$ are minimizers.
Equality at their midpoint forces equality in the convexity inequalities for
the quadratic term and the Euclidean norm. Thus
$(\bm b_1-\bm b_2)^\top\bm M(\bm b_1-\bm b_2)=0$, so
$\bm b_1-\bm b_2\in\ker(\bm M)$, while $\bm b_1$ and $\bm b_2$ lie on the
same nonnegative ray. If they were distinct, every nonzero member of
$\{\bm b_1,\bm b_2\}$ would belong to $\ker(\bm M)$. But for any nonzero
$\bm b\in\ker(\bm M)$, $\bm z^\top\bm b=0$ and hence
$R(\bm b)+\tau\|\bm b\|_2>0=\Phi(\bm0)$, a contradiction. The minimizer is
therefore unique.

For each $a>0$, $\bm M+a\bm I_m\succ0$, so the inner problem is strongly
convex and has a unique minimizer $\bm b(a)$. If the block minimizer
$\widehat{\bm b}$ is nonzero, its KKT condition is
\[
\bm0\in \bm M\widehat{\bm b}-\bm z+\partial R(\widehat{\bm b})
+\tau\widehat{\bm b}/\|\widehat{\bm b}\|_2.
\]
Setting $a=\tau/\|\widehat{\bm b}\|_2$ turns this into the inner optimality
condition, so $\widehat{\bm b}=\bm b(a)$ and
$a\|\widehat{\bm b}\|_2=\tau$. Conversely, these two relations turn the
inner optimality condition back into the KKT condition of the original block
problem. To verify the remaining direction for an arbitrary $a>0$,
suppose $\bm b(a)=\widehat{\bm b}\ne\bm0$. Choose
$\bm q_a,\bm q_*\in\partial R(\widehat{\bm b})$ satisfying the inner
and original KKT conditions, respectively. Subtracting those conditions
and taking the inner product with $\widehat{\bm b}$ gives
\[
(\bm q_a-\bm q_*)^\top\widehat{\bm b}
+a\|\widehat{\bm b}\|_2^2-\tau\|\widehat{\bm b}\|_2=0.
\]
Positive homogeneity implies
$\bm q_a^\top\widehat{\bm b}=\bm q_*^\top\widehat{\bm b}
=R(\widehat{\bm b})$. Since $\widehat{\bm b}\ne\bm0$, division by
its norm yields $a\|\widehat{\bm b}\|_2=\tau$. This completes the equivalence.

\subsection{Proof of Proposition~\ref{main-prop:isotropic} }
\label{supsec:prop1}

We first record three consequences of convex positive homogeneity. If
$\bm q\in\partial R(\bm v)$, then $\bm q^\top\bm v=R(\bm v)$; consequently
$\partial R(t\bm v)=\partial R(\bm v)$ for every $t>0$, and
$\partial R(\bm v)\subseteq\partial R(\bm0)$.

Suppose first that $\bm M=c\bm I_m$. With the change of variables
$\bm v=(c+a)\bm b$, positive homogeneity transforms the inner objective,
up to multiplication by the positive constant $c+a$, into
$\|\bm v\|_2^2/2-\bm z^\top\bm v+R(\bm v)$. Hence
\[
\bm b(a)=(c+a)^{-1}\operatorname{prox}_R(\bm z).
\]
Let $\bm v=\operatorname{prox}_R(\bm z)$. If $\|\bm v\|_2\le\tau$, proximal
optimality gives $\bm z-\bm v\in\partial R(\bm v)\subseteq\partial R(\bm0)$,
so the zero-block KKT condition holds. If $\|\bm v\|_2>\tau$, choose
$a^*=c\tau/(\|\bm v\|_2-\tau)$. Then
$a^*\|\bm b(a^*)\|_2=\tau$, and the reduction lemma yields
\[
\widehat{\bm b}=\frac1c\left(1-\frac{\tau}{\|\bm v\|_2}\right)\bm v.
\]

This establishes (i)$\Rightarrow$(ii) and (i)$\Rightarrow$(iii).

To prove (ii)$\Rightarrow$(i), fix $\bm v\ne\bm0$ and choose
$\bm q\in\partial R(\bm v)$, which is possible because $R$ is finite
and convex. Set $\bm z=\bm v+\bm q$, so
$\operatorname{prox}_R(\bm z)=\bm v$. For any $a>0$, part (ii)
gives $\bm b(a)=\bm v/(c+a)$. Its optimality condition supplies
$\bm q_a\in\partial R(\bm v)$ such that
$(\bm M+a\bm I)\bm v/(c+a)-\bm v-\bm q+\bm q_a=\bm0$.
Taking the inner product with $\bm v$ cancels the two subgradient
terms, since both equal $R(\bm v)$. Hence
$\bm v^\top\bm M\bm v=c\|\bm v\|_2^2$ for every $\bm v$,
and symmetry gives $\bm M=c\bm I$.

For the converse, assume the nested radial formula in part (iii) holds for
every $\bm z$ and $\tau$. Fix $\bm v\ne\bm0$, choose
$\bm q\in\partial R(\bm v)$, and set $\bm z=\bm v+\bm q$, so that
$\operatorname{prox}_R(\bm z)=\bm v$. For any
$0<\tau<\|\bm v\|_2$, the assumed formula gives
$\widehat{\bm b}=\alpha\bm v$ with
$\alpha=c^{-1}(1-\tau/\|\bm v\|_2)>0$. The active KKT condition and
$\partial R(\alpha\bm v)=\partial R(\bm v)$ imply, after taking the inner
product with $\bm v$, that
$\alpha\bm v^\top\bm M\bm v=\|\bm v\|_2(\|\bm v\|_2-\tau)$. Substituting
$\alpha$ gives $\bm v^\top\bm M\bm v=c\|\bm v\|_2^2$. Since $\bm v$ is
arbitrary and $\bm M$ is symmetric, $\bm M=c\bm I_m$. This proves
Proposition~\ref{main-prop:isotropic}.

\subsection{The Yosida lemma used in Theorem~\ref{main-thm:norm_equation}}
\label{supsec:yosida_lemma}

Lemma~\ref{main-lem:yosida} of the main paper is the standard parameter-monotonicity and
minimum-section property of the Yosida approximation for a maximal monotone
subdifferential. For completeness, let $A=\partial g$, with $g$ proper,
closed and convex, and suppose $\bm0\in\operatorname{dom}(A)$.
The resolvent $J_{\lambda A}$ is single valued and firmly nonexpansive,
while $A_\lambda(\bm0)=-J_{\lambda A}(\bm0)/\lambda$.
The classical Yosida theory implies that
$\|A_\lambda(\bm0)\|_2$ is continuous and nonincreasing in
$\lambda>0$ and converges to the norm of the minimum-norm element of
$A(\bm0)$ as $\lambda\downarrow0$; see
\citet{yosida1948differentiability,brezis1973operateurs}.
The application to the present problem, including the endpoint at
$a\downarrow0$, is proved in the next subsection. This standard operator
lemma is not a new monotonicity claim for individual coordinates of
$\bm b(a)$.

\subsection{Proof of Theorem~\ref{main-thm:norm_equation} via Yosida regularization}
\label{supsec:yosida_proof}

Let $g(\bm b)=\bm b^\top\bm M\bm b/2-\bm z^\top\bm b+R(\bm b)$ and
$A=\partial g$. Since $g$ is proper, closed and convex, $A$ is maximal
monotone. The inner optimality condition is
$\bm0\in A\{\bm b(a)\}+a\bm b(a)$. With $\lambda=1/a$,
$\bm b(a)=J_{\lambda A}(\bm0)$ and
$A_\lambda(\bm0)=-a\bm b(a)$, so
$\psi(a)=\|A_{1/a}(\bm0)\|_2$. The classical parameter monotonicity and
continuity of the Yosida approximation therefore imply that $\psi$ is
continuous and nondecreasing on $(0,\infty)$.

For the left endpoint, optimality relative to $\bm0$ and $R\ge0$ give
\[
\frac12\bm b(a)^\top\bm M\bm b(a)-\bm z^\top\bm b(a)
+\frac a2\|\bm b(a)\|_2^2\le0.
\]
Since $\bm z\in\operatorname{range}(\bm M)$,
$\bm b^\top\bm M\bm b/2-\bm z^\top\bm b\ge
-\bm z^\top\bm M^+\bm z/2$. Hence
$\psi(a)^2\le a\bm z^\top\bm M^+\bm z\to0$ as $a\downarrow0$.

For the right endpoint, $R(\bm0)=0$ and $R\ge0$ imply
$\bm0\in\partial R(\bm0)$, so $\bm0\in\operatorname{dom}(A)$. The
minimum-section limit gives $A_{1/a}(\bm0)\to A^0(\bm0)$ as $a\to\infty$.
Since $A(\bm0)=-\bm z+\partial R(\bm0)$,
\[
\lim_{a\to\infty}\psi(a)
=\operatorname{dist}\{\bm z,\partial R(\bm0)\}.
\]
Finally, the zero-block KKT condition is
$\bm0\in-\bm z+\partial R(\bm0)+\tau\mathbb B_2$, equivalent to
$\operatorname{dist}\{\bm z,\partial R(\bm0)\}\le\tau$. If the inequality
is strict in the opposite direction, continuity and the endpoint limits give
a positive root of $\psi(a)=\tau$. The reduction lemma shows that every positive root
produces the unique active minimizer and must satisfy
$a=\tau/\|\widehat{\bm b}\|_2$, so the root is unique.

\subsection{A direct finite-dimensional proof of Theorem~\ref{main-thm:norm_equation}}
\label{supsec:direct_norm_equation}

For completeness, we give a proof that uses only finite-dimensional convex
optimality. Write $\bm b_i=\bm b(a_i)$ and $r_i=\|\bm b_i\|_2$ for
$0<a_1<a_2$. Optimality of the two inner problems gives
\[
g(\bm b_1)+\frac{a_1}{2}r_1^2\le g(\bm b_2)+\frac{a_1}{2}r_2^2,
\qquad
g(\bm b_2)+\frac{a_2}{2}r_2^2\le g(\bm b_1)+\frac{a_2}{2}r_1^2,
\]
so $r_1\ge r_2$. On the other hand,
$-a_i\bm b_i\in\partial g(\bm b_i)$, and monotonicity of $\partial g$
gives
$(r_1-r_2)(a_1r_1-a_2r_2)\le0$. If $r_1>r_2$, division by
$r_1-r_2$ gives the desired inequality. If $r_1=r_2$, the same inequality
follows directly from $a_1<a_2$ and $r_1\ge0$. Thus
$a_1r_1\le a_2r_2$, proving that $\psi(a)=a\|\bm b(a)\|_2$ is
nondecreasing.

To prove continuity, let $a_k\to a>0$. Optimality relative to
$\bm0$ implies
\[
\frac{a_k}{2}\|\bm b(a_k)\|_2^2
\le\|\bm z\|_2\|\bm b(a_k)\|_2,
\]
so the sequence is bounded. By closedness of the graph of $\partial R$,
any convergent subsequence has a limit $\bar{\bm b}$ satisfying
$\bm z-(\bm M+a\bm I_m)\bar{\bm b}\in\partial R(\bar{\bm b})$.
Strong convexity then gives $\bar{\bm b}=\bm b(a)$, proving continuity.

The left endpoint follows from the same Moore--Penrose bound used above. For
the right endpoint, choose $\bm q(a)\in\partial R\{\bm b(a)\}$ satisfying
$\bm M\bm b(a)-\bm z+\bm q(a)+a\bm b(a)=\bm0$. For any
$\bm q_0\in\partial R(\bm0)$, monotonicity of $\partial R$ yields
$a\|\bm b(a)\|_2\le\|\bm z-\bm q_0\|_2$. Hence
$\psi(a)\le d_0:=\operatorname{dist}\{\bm z,\partial R(\bm0)\}$. Setting
$\bm u(a)=a\bm b(a)$ shows that $\bm u(a)$ is bounded and
$\bm b(a)\to\bm0$. Because $R$ is finite, convex and positively
homogeneous, $\partial R(\bm0)$ is compact, and
$\bm q(a)\in\partial R(\bm b(a))\subseteq\partial R(\bm0)$.
Thus the selected subgradients $\bm q(a)$ are bounded. Along any
sequence $a_k\to\infty$, take a subsequence
such that $\bm u(a_k)\to\bar{\bm u}$ and
$\bm q(a_k)\to\bar{\bm q}$. Closedness of $\partial R$ gives
$\bar{\bm q}\in\partial R(\bm0)$, while the inner KKT equation gives
$\bar{\bm u}+\bar{\bm q}=\bm z$. Thus
$\|\bar{\bm u}\|_2\ge d_0$, whereas the preceding upper bound gives the
reverse inequality. Therefore $\psi(a)\to d_0$. The zero--active transition
and uniqueness of the positive root now follow exactly as in the preceding
Yosida proof.

\section{Proof of the certified block bound and valid inner subgradients}
\label{supsec:inner_certificates}

\subsection{Proof of Theorem~\ref{main-thm:certified}}
\label{supsec:proof_certified}

Fix $a>0$ and choose $\bm q^*\in\partial R\{\bm b(a)\}$ so that
$(\bm M+a\bm I_m)\bm b(a)-\bm z+\bm q^*=\bm0$. With
$\tilde{\bm q}\in\partial R(\tilde{\bm b})$ and residual $\bm r_a$ from the
main paper, monotonicity of $\partial R$ and
$\bm M+a\bm I_m\succeq\mu_a\bm I_m$ give
\[
\langle\bm r_a,\tilde{\bm b}-\bm b(a)\rangle
\ge\mu_a\|\tilde{\bm b}-\bm b(a)\|_2^2.
\]
Cauchy--Schwarz therefore yields
$\|\tilde{\bm b}-\bm b(a)\|_2\le\delta_a/\mu_a=e_a$. The reverse triangle
inequality then gives $L_a\le\psi(a)\le U_a$. Since the
positive-part truncation cannot increase the interval width and
$2ae_a\le2\delta_a$, we also have $U_a-L_a\le2\delta_a$.

For $\tilde{\bm b}\ne\bm0$, use the particular subgradient
$\tilde{\bm q}$ in the definition of the original block KKT residual. Since
\[
\bm M\tilde{\bm b}-\bm z+\tilde{\bm q}
+\tau\frac{\tilde{\bm b}}{\|\tilde{\bm b}\|_2}
=\bm r_a+
\left(\frac{\tau}{\|\tilde{\bm b}\|_2}-a\right)\tilde{\bm b},
\]
the triangle inequality gives
$\mathcal R_{\rm blk}(\tilde{\bm b})\le
\delta_a+|a\|\tilde{\bm b}\|_2-\tau|$. If
$L_a\le\tau\le U_a$, then
$|a\|\tilde{\bm b}\|_2-\tau|\le ae_a=(a/\mu_a)\delta_a\le\delta_a$ because
$\mu_a\ge a$. Hence
$\mathcal R_{\rm blk}(\tilde{\bm b})\le2\delta_a$.

\subsection{Convergence of inner iterates and KKT residuals}
\label{supsec:residual_counterexample}

Convergence of the inner iterates or objective values does not necessarily
imply convergence of the corresponding KKT residuals. To see this,
consider the strongly convex function
\[
f(x)=\frac12x^2+|x|,\qquad x\in\mathbb R.
\]
Its unique minimizer is $x^*=0$. For the sequence $x_k=1/k$, we have
$x_k\to x^*$ and
\[
f(x_k)-f(x^*)=\frac{1}{2k^2}+\frac1k\longrightarrow0.
\]
However, since $x_k>0$,
\[
\partial f(x_k)=\left\{1+\frac1k\right\},
\qquad
\operatorname{dist}\{0,\partial f(x_k)\}
=1+\frac1k\longrightarrow1.
\]
Thus convergence of the iterates and objective values, even for a strongly
convex problem, does not ensure a vanishing KKT residual. The example is
also an instance of the inner problem with $m=1$, $M=0$, $a=1$, $z=0$
and $R(x)=|x|$.

This shows why the bounds in Theorem~\ref{main-thm:certified}
require an explicitly evaluated inner residual using a valid subgradient
at the returned iterate. The next subsection shows how to construct such
a subgradient from an exact proximal-gradient step.

\subsection{A valid subgradient from an exact proximal-gradient step}
\label{supsec:pg_subgradient}

Fix $a>0$ and write
\[
\bm Q_a=\bm M+a\bm I_m,\qquad
\mu_a=\lambda_{\min}(\bm M)+a,\qquad
\ell_a=\lambda_{\max}(\bm M)+a.
\]
Thus $\bm Q_a$ is positive definite, with eigenvalues in
$[\mu_a,\ell_a]$. Let $0<\eta_a\le\ell_a^{-1}$ be a proximal-gradient
step size. Starting from any $\bm b^k$, perform the exact update
\begin{equation}
\label{sup:eq:pg_inner_step}
\bm b^{k+1}
=
\operatorname{prox}_{\eta_a R}
\left\{\bm b^k-\eta_a(\bm Q_a\bm b^k-\bm z)\right\}.
\end{equation}
The optimality condition of this proximal step is
\[
\bm0\in
\frac{\bm b^{k+1}-\bm b^k}{\eta_a}
+\bm Q_a\bm b^k-\bm z
+\partial R(\bm b^{k+1}).
\]
Consequently, the vector
\begin{equation}
\label{sup:eq:pg_compatible_subgrad}
\bm q^{k+1}
=
\frac{\bm b^k-\bm b^{k+1}}{\eta_a}
-\bm Q_a\bm b^k+\bm z
\end{equation}
satisfies
\[
\bm q^{k+1}\in\partial R(\bm b^{k+1}).
\]
In particular, validity at the returned iterate follows directly from
the proximal optimality condition; no separate search over the
subdifferential is needed.

The corresponding inner residual at the returned point is
\begin{align}
\bm r_a^{k+1}
&=\bm Q_a\bm b^{k+1}-\bm z+\bm q^{k+1}\notag\\
&=(\bm I_m-\eta_a\bm Q_a)
\frac{\bm b^k-\bm b^{k+1}}{\eta_a}.
\label{sup:eq:pg_inner_residual}
\end{align}
Define the proximal-gradient mapping
\[
\bm G_{\eta_a}(\bm b^k)
=
\frac{\bm b^k-\bm b^{k+1}}{\eta_a}.
\]
Since $0<\eta_a\le\ell_a^{-1}$, all eigenvalues of
$\bm I_m-\eta_a\bm Q_a$ lie in
$[0,1-\eta_a\mu_a]$. Therefore
\begin{equation}
\label{sup:eq:pg_residual_bound}
\delta_a^{k+1}
:=\|\bm r_a^{k+1}\|_2
\le
(1-\eta_a\mu_a)
\|\bm G_{\eta_a}(\bm b^k)\|_2.
\end{equation}
Thus every exact proximal-gradient step produces a subgradient at the
current returned point that is admissible in the residual definition
of Theorem~\ref{main-thm:certified}. As the proximal-gradient iterates
converge to the unique minimizer of the strongly convex inner problem,
the mapping and the valid inner residual converge to zero. This argument
applies to any proper closed convex $R$ for which the proximal step can
be evaluated exactly. If the proximal operator is itself evaluated
inexactly, its optimality residual must also be controlled; the
subgradient formula above need not be valid at an inexact proximal output.

\section{Proofs for sweep-level KKT control and estimation and prediction error bounds}
\label{supsec:global_proofs}

\subsection{Proof of Theorem~\ref{main-thm:sweep_global}}
\label{supsec:proof_sweep}

Let $\varphi_j(\bm b)=R_j(\bm b)+\tau_j\|\bm b\|_2$,
so that $F_n=f_n+\sum_j\varphi_j$. After updating block $j$ in
sweep $t$, choose $\bm p_j^{(t)}\in
\partial\varphi_j\{\bm\beta_j^{(t,j)}\}$ and write the certified
local first-order residual as
\[
\bm g_{j,\mathrm{loc}}^{(t)}
=\nabla_j f_n\{\bm\beta^{(t,j)}\}+\bm p_j^{(t)},
\qquad
\|\bm g_{j,\mathrm{loc}}^{(t)}\|_2
\le\varepsilon_j^{(t)}.
\]
Since the block coefficient does not change after its update,
$\bm p_j^{(t)}$ is still a valid subgradient of
$\varphi_j$ at $\bm\beta_j^{(t+1)}$. For squared loss,
\[
\nabla_j f_n\{\bm\beta^{(t+1)}\}
-\nabla_j f_n\{\bm\beta^{(t,j)}\}
=\sum_{k=j+1}^G\frac{\bm X_j^\top\bm X_k}{n}
(\bm\beta_k^{(t+1)}-\bm\beta_k^{(t)})
=\bm s_j^{(t)}.
\]
Therefore
\[
\bm g_j^{(t+1)}:=\bm g_{j,\mathrm{loc}}^{(t)}
+\bm s_j^{(t)}
\in\nabla_j f_n\{\bm\beta^{(t+1)}\}
+\partial\varphi_j\{\bm\beta_j^{(t+1)}\},
\]
and $\|\bm g_j^{(t+1)}\|_2\le
\varepsilon_j^{(t)}+\|\bm s_j^{(t)}\|_2$.
Concatenating the blocks gives the global subgradient asserted in
Theorem~\ref{main-thm:sweep_global}.

For the statistical normalization used in the main paper, let
$\varphi_j=\lambda w_j\omega_j$, with
$w_j=\tau_j/\lambda$ and
$\omega_j(\bm b)=\|\bm b\|_2+R_j(\bm b)/\tau_j$.
The dual of the block-separable norm
$\omega(\bm\beta)=\sum_jw_j\omega_j(\bm\beta_j)$ is
\[
\omega^*(\bm g)=\max_j
\frac{\omega_j^*(\bm g_j)}{w_j}.
\]
Since $\omega_j(\bm b)\ge\|\bm b\|_2$,
$\omega_j^*(\bm u)\le\|\bm u\|_2$.
Thus the constructed subgradient satisfies
\[
\omega^*\{\bm g^{(t+1)}\}
\le\max_j\frac{\varepsilon_j^{(t)}+
\|\bm s_j^{(t)}\|_2}{w_j}
=\delta_{\rm sw}^{(t)},
\]
which also proves the sweep-level bound used in Section~3.3.

\subsection{Proof of Theorem~\ref{main-thm:statistical_consistency}}
\label{supsec:statistical_proof}
The norm is decomposable across the selected whole groups because it is a
sum of block norms. No decomposability within a fused block is assumed.
The proof applies to any approximate estimator admitting the stated
global KKT bound, independently of the optimization algorithm.

Let $\widetilde{\bm\beta}$ be such an estimator, set
$\bm h=\widetilde{\bm\beta}-\bm\beta^*$ and
$\eta=\omega^*(\bm X^\top\bm\varepsilon/n)$, and suppose
there exists $\bm g\in\partial F_n(\widetilde{\bm\beta})$
with $\omega^*(\bm g)\le\delta$. For the norm-equation
algorithm, Theorem~\ref{main-thm:sweep_global} supplies this bound
with $\widetilde{\bm\beta}=\bm\beta^{(t+1)}$ and
$\delta=\delta_{\rm sw}^{(t)}$.
The subgradient inequality and the model
$\bm y=\bm X\bm\beta^*+\bm\varepsilon$ give
\begin{equation}
\label{sup:eq:approx_basic}
\frac{\|\bm X\bm h\|_2^2}{2n}
\le(\eta+\delta)\omega(\bm h)
+\lambda\{\omega(\bm\beta^*)-
\omega(\bm\beta^*+\bm h)\}.
\end{equation}
Write $S=\{j:\bm\beta_j^*\ne\bm0\}$,
let $\mathcal M$ be the group-support subspace and
$\mathcal M^\perp$ its complement. By decomposability,
\[
\omega(\bm\beta^*)-
\omega(\bm\beta^*+\bm h)
\le\omega(\bm h_{\mathcal M})
-\omega(\bm h_{\mathcal M^\perp}).
\]
With $\theta=\gamma+\xi<1$ and
$\eta+\delta\le\theta\lambda$, we obtain
\begin{equation}
\label{sup:eq:cone_basic}
\frac{\|\bm X\bm h\|_2^2}{2n}
+\lambda(1-\theta)\omega(\bm h_{\mathcal M^\perp})
\le\lambda(1+\theta)\omega(\bm h_{\mathcal M}).
\end{equation}
Nonnegativity of the squared-loss term gives the cone condition
$\omega(\bm h_{\mathcal M^\perp})\le
C_\theta\omega(\bm h_{\mathcal M})$,
where $C_\theta=(1+\theta)/(1-\theta)$.
On that cone, Assumption~\ref{main-ass:restricted_eigenvalue} of the main paper yields
$\|\bm X\bm h\|_2^2/n\ge\kappa\|\bm h\|_2^2$.
Using the subspace compatibility constant
$\omega(\bm h_{\mathcal M})\le
\Psi(\mathcal M)\|\bm h\|_2$ in
\eqref{sup:eq:cone_basic} gives
\[
\|\bm h\|_2
\le\frac{2(1+\theta)\lambda\Psi(\mathcal M)}{\kappa}.
\]
Dropping the nonpositive complement term in
\eqref{sup:eq:approx_basic} and substituting the latter bound gives
\[
\frac{\|\bm X\bm h\|_2^2}{n}
\le\frac{4(1+\theta)^2\lambda^2\Psi^2(\mathcal M)}{\kappa}.
\]
These are the two asserted bounds. If the score control and restricted
curvature events have probability tending to one,
$\kappa$ stays bounded away from zero and
$\lambda_n\Psi(\mathcal M_n)\to0$, the same bounds yield the
orders in the main paper and convergence in probability.

\subsection{Proof of the group-lasso corollary}
\label{supsec:group_corollary}

For the ordinary group lasso with equal group size $m$ and weights
$w_j=\sqrt m$, the normalized block norm is
$\omega_j(\bm b)=\|\bm b\|_2$ and
$\Psi(\mathcal M)=\sqrt{sm}$ when $s$ groups are active.
Conditionally on $\bm X$, define
$\bm Z_j=\bm X_j^\top\bm\varepsilon/(n\sqrt m)$.
Under Gaussian errors and the design bound
$\lambda_{\max}(\bm X_j^\top\bm X_j/n)\le C_X$,
$\bm Z_j$ is dominated in Euclidean norm by
$\sigma\sqrt{C_X/(nm)}\|\bm Z\|_2$,
where $\bm Z\sim N(\bm0,\bm I_m)$.
The Gaussian concentration inequality
\[
\Pr\{\|\bm Z\|_2\ge\sqrt m+\sqrt{2x}\}
\le e^{-x}
\]
and a union bound with $x=\log(G/\alpha)$ show that, with
probability at least $1-\alpha$,
\[
\eta\le\frac{\sigma\sqrt{C_X}}{\sqrt n}
\left\{1+\sqrt{\frac{2\log(G/\alpha)}{m}}\right\}.
\]
The main paper's choice of $\lambda_n$ ensures
$\eta\le\gamma\lambda_n$. Applying Theorem~\ref{main-thm:statistical_consistency} with
$\Psi(\mathcal M)=\sqrt{sm}$ proves the corollary.

\section{Algorithms and implementation details}
\label{supsec:implementation}

\subsection{The common block problem and the compared methods}
\label{supsec:methods_overview}

Across the main and supplementary experiments, we consider three
accuracy strategies for the norm-equation (NormEq) algorithm. The main
sparse-group and fused sparse-group comparisons also include direct
proximal-gradient methods. An additional sparse-group experiment
compares NormEq with the signed single-line search (SSLS) benchmark.
This section specifies the common block problem, the SSLS
configuration search, the certified NormEq acceptance rule, the inner
solvers, the full-sweep stopping criteria and the evaluation measures.
The theoretical results do not depend on these particular numerical
implementations.

At a cyclic update of block $j$, write
\[
\bm r_j=\bm y-\sum_{k\ne j}\bm X_k\bm\beta_k,\quad
\bm M_j=\bm X_j^\top\bm X_j/n,\quad
\bm z_j=\bm X_j^\top\bm r_j/n.
\]
Suppressing $j$, the block problem is
\begin{equation}
\label{sup:eq:common_block}
\min_{\bm b}\left\{\frac12\bm b^\top\bm M\bm b-
\bm z^\top\bm b+R(\bm b)+\tau\|\bm b\|_2\right\}.
\end{equation}
For the sparse-group lasso, $R(\bm b)=\lambda_1\|\bm b\|_1$ and
$\tau=\lambda_2\sqrt m$. For the fused sparse-group model,
$R(\bm b)=\lambda_1\|\bm b\|_1+
\lambda_f\|\bm D\bm b\|_1$ and $\tau=\lambda_g\sqrt m$.

The three NormEq versions share the same scalar characterization and zero
block test. \emph{NormEq--Exact} uses high-accuracy inner solves and a
high-accuracy scalar search. \emph{NormEq--Joint} instead stops each inner
solve when a scalar-search direction is certified or when the combined
inner--outer block KKT bound meets a block tolerance; the completed BCD fit
still targets high global accuracy. \emph{NormEq--Early} uses the same
certified block mechanism but terminates BCD at a sweep-level
regularization-scaled threshold, with a local block tolerance
specified for the particular experiment.

The SSLS benchmark uses the same support/sign configurations,
conditional group-lasso equation and KKT compatibility tests as the
original method. We use the safeguarded Newton solver described below.
Bisection and an adaptive-start Newton variant were used in a preliminary
root-solver comparison; the latter's additional function evaluations
outweighed its reduction in Newton iterations, so SSLS--Newton is retained
as the representative implementation.

\subsection{SSLS: conditional equation and valid sign configurations}
\label{supsec:ssls}
The spectral SSLS formulas and finite brackets in this and the next
subsection are stated for $\bm M\succ0$, so every principal-block
eigenvalue $d_i$ is positive. This restriction applies to the benchmark
implementation, not to the NormEq theory. With a singular principal block,
the shifted score need not belong to its range, and the formulas involving
$v_i/d_i$ cannot be used without a separate null-space analysis.

For $R(\bm b)=\lambda_1\|\bm b\|_1$, first test the exact zero-block
condition
\begin{equation}
\label{sup:eq:ssls_zero}
\bigl\|S_{\lambda_1}(\bm z)\bigr\|_2\le\tau,
\end{equation}
where $S_{\lambda_1}$ is coordinatewise soft thresholding. If this holds,
return $\bm0$. Otherwise, take a candidate
$\bm s\in\{-1,0,1\}^m$ and $J=\{i:s_i\ne0\}$. A valid nonzero
solution with the proposed active signs must satisfy
\[
\bm M_{JJ}\bm b_J-(\bm z_J-\lambda_1\bm s_J)
+\tau\frac{\bm b_J}{\|\bm b_J\|_2}=\bm0,
\qquad \bm b_{J^c}=\bm0.
\]
Put $\bm q_J=\bm z_J-\lambda_1\bm s_J$. If
$\|\bm q_J\|_2\le\tau$, the conditional group-lasso solution is zero
and this nonempty candidate is rejected. Otherwise, for the positive
definite principal block used in the experiments, let
\[
\bm M_{JJ}=\bm Q\operatorname{diag}(d_1,\ldots,d_r)\bm Q^\top,
\quad r=|J|,\quad d_i>0,\quad \bm v=\bm Q^\top\bm q_J.
\]
Writing $h=\|\bm b_J\|_2>0$ gives
\begin{equation}
\label{sup:eq:ssls_phi}
\begin{aligned}
\phi_J(h)&=\sum_{i=1}^r\frac{v_i^2}{(d_i h+\tau)^2}-1=0,\\
\phi_J'(h)&=-2\sum_{i=1}^r\frac{d_i v_i^2}{(d_i h+\tau)^3},\qquad
\phi_J''(h)=6\sum_{i=1}^r\frac{d_i^2v_i^2}{(d_i h+\tau)^4}.
\end{aligned}
\end{equation}
The function is strictly decreasing and convex on $h\ge0$, with
$\phi_J(0)>0$. Its positive root determines the conditional coefficients
\begin{equation}
\label{sup:eq:ssls_recover}
\bm b_J=h\bm Q
\operatorname{diag}\{(d_i h+\tau)^{-1}\}_{i=1}^r\bm v,
\qquad \bm b_{J^c}=\bm0.
\end{equation}
A candidate is accepted only if the recovered $\bm b_J$ has the proposed
strict signs and the inactive-coordinate condition holds:
\begin{equation}
\label{sup:eq:ssls_compatibility}
\operatorname{sign}(b_i)=s_i\quad(i\in J),\qquad
\bigl|z_i-\bm M_{iJ}\bm b_J\bigr|\le\lambda_1
\quad(i\notin J).
\end{equation}
The active-coordinate KKT residual is also checked numerically. Since each
coordinate has three possible states, there are at most $3^m$ possible
sign/support configurations \emph{per block update}. A full BCD fit may
check the same configuration at more than one block update; its cumulative
configuration count need not be bounded by $3^m$.

\subsection{Newton solver used for SSLS}
\label{supsec:ssls_roots}

For a proposed active set $J$, a valid upper bracket for the positive
root of \eqref{sup:eq:ssls_phi} is
\begin{equation}
\label{sup:eq:ssls_upper}
 h_{\rm U}=\left\|\left(v_i/d_i\right)_{i=1}^r\right\|_2,
 \qquad \phi_J(h_{\rm U})\le0.
\end{equation}
We also use the sharper lower bound
\begin{equation}
\label{sup:eq:ssls_lower}
 h_{\rm L}=\max\left\{0,
 \frac{-\tau D_1+
 \sqrt{\{\tau D_1\}^2-D_2(r\tau^2-\|\bm v\|_1^2)}}{D_2}
 \right\},
\end{equation}
when the square root is real, where
$D_1=\sum_i d_i$ and $D_2=\sum_i d_i^2$; otherwise set
$h_{\rm L}=0$. We verify the proposed lower bound by evaluating
$\phi_J(h_{\rm L})$. If it is not a valid lower bracket, the
implementation reverts to zero.

Starting from the verified lower endpoint, SSLS--Newton iterates
\begin{equation}
\label{sup:eq:ssls_newton}
 h_{k+1}=h_k-\frac{\phi_J(h_k)}{\phi_J'(h_k)}.
\end{equation}
Because $\phi_J$ is strictly decreasing and convex, Newton iterations
started at a valid lower bracket approach the root from below. We retain
the bracket as a safeguard and use its midpoint if a proposed Newton
step is nonfinite or leaves the current bracket. The iteration stops
when the scalar residual or bracket width reaches the prescribed
root tolerance.

In the preliminary comparison at $n=200$, $G=20$ and $m=5$,
SSLS--Newton required a mean of 17,389.7 root evaluations and took
0.2823 seconds per fit, compared with 77,966.2 evaluations and
0.5247 seconds for bisection. Newton--ABS required 26,843.2
evaluations and took 0.3210 seconds: its adaptive initialization
did not offset the additional function evaluations. We therefore
use SSLS--Newton in the comparison and scaling experiments below.

\begin{algorithm}[H]
\caption{SSLS block update with safeguarded Newton iterations}
\label{sup:alg:ssls}
\begin{algorithmic}[1]
\REQUIRE $\bm M,\bm z,\lambda_1,\tau$; root tolerance
$\varepsilon_{\rm root}$
\IF{$\|S_{\lambda_1}(\bm z)\|_2\le\tau$}
  \STATE \textbf{return} $\bm0$
\ENDIF
\FOR{candidate $\bm s\in\{-1,0,1\}^m$ in the common search order}
 \STATE Set $J=\{i:s_i\ne0\}$; skip $J=\varnothing$
 \STATE Form $\bm q_J=\bm z_J-\lambda_1\bm s_J$;
 skip if $\|\bm q_J\|_2\le\tau$
 \STATE Form or reuse
 $\bm M_{JJ}=\bm Q\operatorname{diag}(d_i)\bm Q^\top$
 \STATE Construct and verify $[h_{\rm L},h_{\rm U}]$
 \STATE Solve \eqref{sup:eq:ssls_phi} by safeguarded Newton;
 recover $\bm b$ by \eqref{sup:eq:ssls_recover}
 \IF{the sign and inactive-coordinate tests
 \eqref{sup:eq:ssls_compatibility} hold}
  \STATE \textbf{return} $\bm b$
 \ENDIF
\ENDFOR
\STATE Report failure if no candidate passes the KKT checks
\end{algorithmic}
\end{algorithm}

The experiments use a common deterministic candidate order, testing the
previous block sign vector and the soft-thresholded score sign before an
exhaustive lexicographic fallback. Principal-block eigendecompositions are
cached by active index set; only the shifted score changes with the sign
configuration. A candidate is counted each time it is visited, including
repeated visits in later sweeps.

\subsection{Certified NormEq block updates: inner and outer accuracy}
\label{supsec:certified_algorithm}

For arbitrary convex positively homogeneous $R$, the exact
zero-block condition is
\[
\operatorname{dist}\{\bm z,\partial R(\bm0)\}\le\tau.
\]
For an active block, the exact inner problem and outer equation are,
respectively,
\begin{equation}
\label{sup:eq:normeq_inner_outer}
\begin{aligned}
\bm b(a)&=\argmin_{\bm b}\bigl\{
\tfrac12\bm b^\top(\bm M+a\bm I_m)\bm b\\
&\hspace{3em}-\bm z^\top\bm b+R(\bm b)\bigr\},\\
a\|\bm b(a)\|_2&=\tau.
\end{aligned}
\end{equation}
For a trial $a>0$, compute an approximate $\widetilde{\bm b}_a$ and a
\emph{valid} subgradient at that iterate,
$\widetilde{\bm q}_a\in\partial R(\widetilde{\bm b}_a)$.
Define
\begin{equation}
\label{sup:eq:normeq_certificates}
\begin{aligned}
\delta_a&=\|(\bm M+a\bm I_m)\widetilde{\bm b}_a
-\bm z+\widetilde{\bm q}_a\|_2,&
\nu_a&=|a\|\widetilde{\bm b}_a\|_2-\tau|,\\
 e_a&=\frac{\delta_a}{\lambda_{\min}(\bm M)+a},&
L_a&=a(\|\widetilde{\bm b}_a\|_2-e_a)_+,\quad
U_a=a(\|\widetilde{\bm b}_a\|_2+e_a).
\end{aligned}
\end{equation}
Theorem~\ref{main-thm:certified} gives two \emph{different} uses of these quantities. First,
$U_a<\tau$ certifies $a<a^*$, whereas $L_a>\tau$ certifies $a>a^*$;
the inner solve may stop immediately at either decision. Second,
for a nonzero iterate, the joint bound
\begin{equation}
\label{sup:eq:normeq_accept}
\mathcal R_{\rm blk}(\widetilde{\bm b}_a)
\le\delta_a+\nu_a
\end{equation}
allows direct acceptance when
$\delta_a+\nu_a\le\varepsilon_{\rm blk}$, regardless of whether the
inner or scalar equations have been solved to a fixed high precision.
If $L_a\le\tau\le U_a$, then $\nu_a\le\delta_a$ and the convenient
sufficient condition $2\delta_a\le\varepsilon_{\rm blk}$ is available.

An inconclusive interval is refined at the same $a$; the unverified sign
of $a\|\widetilde{\bm b}_a\|_2-\tau$ is never used for bracketing.
Take $\varepsilon_{\rm blk}>0$ and set
$\bar\delta=\min\{\varepsilon_{\rm blk},\tau\}/4$.
Initial expansion and bisection use the same acceptance-first rule.
Start from any $a>0$. If the nonzero iterate meets the block tolerance,
return it. Otherwise, a bracket update is permitted only when
$\delta_a\le\bar\delta$: if $L_a>\tau$, retain $a$ as an upper
endpoint; if $U_a<\tau$, double $a$. In every other case refine the
inner solve at the same trial. Once an upper endpoint is available,
always choose the midpoint of the current bracket. This also handles a trial
at the exact root, where a strict sign certificate need not become
available. The left endpoint can initially be zero by the endpoint limit.

For the termination argument below, assume that at each fixed $a>0$
the inner routine can attain any prescribed positive inner-residual
tolerance, evaluated with a valid subgradient, in finitely many steps.
Strong convexity gives the error bound once a residual is supplied;
it does not prove this property for an arbitrary
inner algorithm. Exhausting an iteration limit must return an unresolved
status, unless the recorded certificate already meets the requested tolerance.

\begin{algorithm}[H]
\caption{Joint-certified NormEq block update}
\label{sup:alg:normeq_block}
\begin{algorithmic}[1]
\REQUIRE $\bm M,\bm z,R,\tau,\varepsilon_{\rm blk}>0$
\STATE Set $\bar\delta=\min\{\varepsilon_{\rm blk},\tau\}/4$
\STATE Check the zero-block condition or a certified zero-block residual; return if accepted
\STATE Expand trials using the acceptance-first rule above; return if accepted
\STATE Otherwise retain $a_{\rm U}$ with $L_{a_{\rm U}}>\tau$ and set $a_{\rm L}=0$
\WHILE{the block has not been accepted}
  \STATE Set $a=(a_{\rm L}+a_{\rm U})/2$
  \REPEAT
   \STATE Perform one or more warm-started steps on the convex inner problem
   \STATE Compute a valid subgradient at the current iterate and the quantities in
   \eqref{sup:eq:normeq_certificates}
   \IF{$\widetilde{\bm b}_a\ne\bm0$ and
   $\delta_a+\nu_a\le\varepsilon_{\rm blk}$}
    \STATE \textbf{return} $\widetilde{\bm b}_a$ and its block KKT bound
   \ELSIF{$\delta_a\le\bar\delta$ and $U_a<\tau$}
    \STATE Set $a_{\rm L}=a$; exit inner refinement at this trial
   \ELSIF{$\delta_a\le\bar\delta$ and $L_a>\tau$}
    \STATE Set $a_{\rm U}=a$; exit inner refinement at this trial
   \ELSE
    \STATE Continue inner refinement at the same $a$
   \ENDIF
  \UNTIL{a bracket endpoint has been updated}
\ENDWHILE
\end{algorithmic}
\end{algorithm}

For a certified active block, the inner-solver assumption ensures a finite
decision at each trial: sufficiently small $\delta_a$ gives acceptance at
$a^*$ and a direction elsewhere. The endpoint limits make expansion finite.
If bisection continues indefinitely, the trials tend to $a^*$, while every
endpoint update satisfies $\delta_a\le\bar\delta$ and
\[
\delta_a+\nu_a
\le 2\delta_a+|\psi(a)-\tau|
\le \varepsilon_{\rm blk}/2+|\psi(a)-\tau|.
\]
At zero, any inner residual formed using a valid subgradient exceeds
$\tau$ in the active-block case, so iterates with
$\delta_a\le\bar\delta$ are nonzero. Continuity then forces the
acceptance test to succeed, a contradiction. This proves finite block acceptance in exact arithmetic
under the stated assumption, without a uniform complexity bound or a
termination guarantee for full cyclic BCD.

The \emph{Exact} implementation uses the same zero test and inner--outer
characterization, but solves each inner problem to a tight residual
tolerance and uses a high-accuracy scalar bracket. It accepts the final
block only after a high-accuracy KKT check. The \emph{Joint} and
\emph{Early} implementations use Algorithm~\ref{sup:alg:normeq_block},
with different block tolerances and full-sweep stopping rules.

\par\noindent\textit{Numerical zero tests.}
Let $C=\partial R(\bm0)$. A feasible $\bm q_0\in C$ gives the
valid zero-block residual bound
\[
\mathcal R_{\rm blk}(\bm0)
=\{\operatorname{dist}(\bm z,C)-\tau\}_+
\le (\|\bm z-\bm q_0\|_2-\tau)_+.
\]
Thus zero can be returned within tolerance even when an exact equality
test is unresolved. To declare a block strictly active using a numerical
projection onto $C$, a lower bound on the distance exceeding $\tau$ is
needed; a feasible projection candidate supplies only an upper bound.
Near the threshold, refine the distance calculation or return zero with
its recorded residual bound. A rounded distance estimate alone does not
certify the exact zero/active decision.

\subsection{Inner solvers and valid subgradients}
\label{supsec:inner_implementation}

For $R(\bm b)=\lambda_1\|\bm b\|_1$, write
$\bm Q_a=\bm M+a\bm I_m$. A coordinate-descent update is
\begin{equation}
\label{sup:eq:sgl_cd}
 b_i\leftarrow\frac{S_{\lambda_1}
 \left(z_i-\sum_{k\ne i}M_{ik}b_k\right)}{M_{ii}+a}.
\end{equation}
We update coordinates in the order $1,\ldots,m$. Let
$\bm b^{(i)}$ denote the vector immediately after coordinate $i$
has been updated, and let $\bm b^{\rm end}$ be the vector after the
complete sweep. The exact one-dimensional minimization at coordinate
$i$ gives a value $q_i^{\rm CD}$ satisfying
\[
q_i^{\rm CD}\in\lambda_1\partial|b_i^{(i)}|,
\qquad
(\bm Q_a\bm b^{(i)}-\bm z)_i+q_i^{\rm CD}=0.
\]
Since coordinate $i$ is not changed again during this sweep,
$b_i^{(i)}=b_i^{\rm end}$. Hence
$q_i^{\rm CD}\in\lambda_1\partial|b_i^{\rm end}|$, and the vector
$\bm q^{\rm CD}=(q_1^{\rm CD},\ldots,q_m^{\rm CD})^\top$ is a
valid subgradient at the final iterate. Its residual has the
explicit form
\begin{equation}
\label{sup:eq:cd_sweep_residual}
\begin{aligned}
(\bm Q_a\bm b^{\rm end}-\bm z+\bm q^{\rm CD})_i
&=(\bm Q_a(\bm b^{\rm end}-\bm b^{(i)}))_i\\
&=\sum_{k>i}M_{ik}(b_k^{\rm end}-b_k^{(i)}).
\end{aligned}
\end{equation}
Thus the residual is exactly the change in coordinate $i$'s
quadratic gradient caused by the later coordinate updates.

This establishes the validity of the CD subgradient at the returned
iterate. We further show that its residual converges to zero under
repeated exact CD sweeps. Write
\[
F_a(\bm b)=\tfrac12\bm b^\top\bm Q_a\bm b
-\bm z^\top\bm b+\lambda_1\|\bm b\|_1,
\]
and index complete sweeps by $s$. Let $\bm b^{(s,i-1)}$ and
$\bm b^{(s,i)}$ be the iterates immediately before and after the
update of coordinate $i$ during sweep $s$, and set
$\Delta_i^{(s)}=b_i^{(s,i)}-b_i^{(s,i-1)}$.
Since the coordinate subproblem is strongly convex with curvature
$(\bm Q_a)_{ii}\ge a$, exact coordinate minimization gives
\[
F_a\{\bm b^{(s,i-1)}\}-F_a\{\bm b^{(s,i)}\}
\ge \frac a2|\Delta_i^{(s)}|^2.
\]
Summing over all coordinates and sweeps, and using that $F_a$ is
bounded below, yields
\[
\sum_{s=0}^{\infty}\sum_{i=1}^m|\Delta_i^{(s)}|^2<\infty.
\]
Hence $\Delta_i^{(s)}\to0$ for every coordinate $i$. Applying
\eqref{sup:eq:cd_sweep_residual} to sweep $s$ gives
\[
 r_{a,i}^{\mathrm{CD},(s)}
 =\sum_{k>i}M_{ik}\Delta_k^{(s)}\longrightarrow0.
\]
As $m$ is finite,
\[
\delta_a^{\mathrm{CD},(s)}
=\|\bm r_a^{\mathrm{CD},(s)}\|_2\longrightarrow0.
\]
Thus exact cyclic CD supplies a valid subgradient after every sweep
and reaches any prescribed positive inner-residual tolerance in
finitely many sweeps. This is an exact-arithmetic statement and does
not give an iteration-complexity bound.

For a given final iterate, an even smaller residual can be obtained
by choosing the minimum-residual admissible $\ell_1$ subgradient.
Let $\bm c=\bm Q_a\bm b^{\rm end}-\bm z$, and write
$\operatorname{clip}(x,l,u)=\max\{l,\min(x,u)\}$ for $l\le u$.
Set
\begin{equation}
\label{sup:eq:sgl_subgrad}
 \widetilde q_i=
 \begin{cases}
 \lambda_1\operatorname{sign}(b_i^{\rm end}),
     &b_i^{\rm end}\ne0,\\
 \operatorname{clip}(-c_i,-\lambda_1,\lambda_1),
     &b_i^{\rm end}=0.
 \end{cases}
\end{equation}
This is a valid element of
$\lambda_1\partial\|\bm b^{\rm end}\|_1$ and minimizes
$\|\bm c+\bm q\|_2$ over all admissible $\ell_1$ subgradients.
The residual $\delta_a=\|\bm c+\widetilde{\bm q}\|_2$ is therefore
no larger than the residual obtained from the coordinatewise
subgradients in \eqref{sup:eq:cd_sweep_residual}.

For the sparse-group lasso, the inner problem can also be solved by proximal gradient. Each exact proximal step yields a valid
subgradient at the returned iterate through
\eqref{sup:eq:pg_compatible_subgrad}; the corresponding residual
bound is given in \eqref{sup:eq:pg_residual_bound}. No separate
subgradient search is required when the proximal operator is
evaluated exactly.

For the fused sparse-group lasso, the inner problem is solved through
a box-constrained dual formulation, followed by a subgradient validity
check at the recovered primal iterate. Specifically, with
$R(\bm b)=\lambda_1\|\bm b\|_1+
\lambda_f\|\bm D\bm b\|_1$, the fixed-$a$ primal inner problem is
\begin{equation}
\label{sup:eq:fused_inner_primal}
 \min_{\bm b}\left\{\tfrac12\bm b^\top\bm Q_a\bm b
 -\bm z^\top\bm b+\lambda_1\|\bm b\|_1
 +\lambda_f\|\bm D\bm b\|_1\right\}.
\end{equation}
We solve its box-constrained dual problem
\begin{equation}
\label{sup:eq:fused_inner_dual}
\min_{\substack{\|\bm u\|_\infty\le\lambda_1\\
\|\bm v\|_\infty\le\lambda_f}}
 \frac12(\bm z-\bm u-\bm D^\top\bm v)^\top
 \bm Q_a^{-1}(\bm z-\bm u-\bm D^\top\bm v),
\end{equation}
and recover the associated primal candidate as
\begin{equation}
\label{sup:eq:fused_primal_recovery}
 \widetilde{\bm b}=\bm Q_a^{-1}
 (\bm z-\bm u-\bm D^\top\bm v).
\end{equation}
The box constraints make $(\bm u,\bm v)$ dual feasible, but do not
by themselves ensure that $\bm u+\bm D^\top\bm v$ is a subgradient
of $R$ at the approximate primal point $\widetilde{\bm b}$.
We therefore construct a valid subgradient at the recovered primal point.
Specifically, impose
\begin{equation}
\label{sup:eq:fused_compatible}
\begin{aligned}
\widetilde q_i^{(1)}
&=\lambda_1\operatorname{sign}(\widetilde b_i)
&&\text{if }\widetilde b_i\ne0,\\
\widetilde q_e^{(f)}
&=\lambda_f\operatorname{sign}
\{(\bm D\widetilde{\bm b})_e\}
&&\text{if }(\bm D\widetilde{\bm b})_e\ne0,
\end{aligned}
\end{equation}
and constrain the remaining coordinates to their respective boxes.
Among these admissible subgradient choices, select one by solving
\begin{equation}
\label{sup:eq:fused_compatible_ls}
\min_{\substack{\bm q^{(1)},\bm q^{(f)}
\text{ satisfy }\eqref{sup:eq:fused_compatible}\\
\|\bm q^{(1)}\|_\infty\le\lambda_1,\ 
\|\bm q^{(f)}\|_\infty\le\lambda_f}}
\frac12\left\|
\bm Q_a\widetilde{\bm b}-\bm z
+\bm q^{(1)}+\bm D^\top\bm q^{(f)}
\right\|_2^2.
\end{equation}
The resulting
$\widetilde{\bm q}=\bm q^{(1)}+\bm D^\top\bm q^{(f)}$
belongs to $\partial R(\widetilde{\bm b})$ by construction.
We use the resulting residual in
\eqref{sup:eq:normeq_certificates} only after explicitly checking
its value against the required tolerance. If it is too large, the
inner dual problem is refined or the candidate face is re-solved;
a small dual objective gap alone is not used as a substitute for
the valid inner optimality residual.

For the zero-block test, the subdifferential is
\[
\partial R(\bm0)=
\{\bm u+\bm D^\top\bm v:
\|\bm u\|_\infty\le\lambda_1,\
\|\bm v\|_\infty\le\lambda_f\},
\]
so the zero condition is checked using the corresponding constrained
Euclidean-distance problem. A feasible pair whose distance to
$\bm z$ is no greater than $\tau$ certifies a zero block; otherwise,
the distance calculation is refined or an independent KKT check
is used.

\subsection{Full cyclic BCD and stopping rules}
\label{supsec:full_bcd}

For each method, the regression solver initializes at
$\bm\beta^{(0)}=\bm0$, visits all groups cyclically, and updates the
partial residual after each block. Once a sweep is complete, the global
Euclidean KKT residual is
\begin{equation}
\label{sup:eq:global_euclidean_kkt}
\mathcal R_{\rm KKT}(\bm\beta)=
\max_j\operatorname{dist}\left\{\bm0,
\nabla_j f_n(\bm\beta)+
\partial\varphi_j(\bm\beta_j)\right\},\quad
\varphi_j=R_j+\tau_j\|\cdot\|_2.
\end{equation}
For high-accuracy \emph{Exact} and \emph{Joint} in the sparse-group
experiments, this quantity is compared to the prescribed KKT tolerance.
The fused sparse-group experiment instead uses the common PG mapping
as its primary numerical target, with a separate KKT check for NormEq.
The \emph{Early} version computes the per-sweep bound
\begin{equation}
\label{sup:eq:full_sweep_bound}
\delta_{\rm sw}^{(t)}=
\max_j\frac{\varepsilon_j^{(t)}+
\|\sum_{k>j}\bm M_{jk}\bm\Delta_k^{(t)}\|_2}{w_j},
\qquad w_j=\frac{\tau_j}{\lambda},
\end{equation}
and stops as soon as
\begin{equation}
\label{sup:eq:full_early_stop}
\delta_{\rm sw}^{(t)}\le\xi\lambda.
\end{equation}
Here $\varepsilon_j^{(t)}$ is a valid local block KKT bound at its
update, and $\bm\Delta_k^{(t)}=\bm\beta_k^{(t+1)}-
\bm\beta_k^{(t)}$. For the sparse-group lasso, $\lambda=\lambda_2$;
for fused sparse-group regression, $\lambda=\lambda_g$.

\begin{algorithm}[H]
\caption{Cyclic NormEq with three accuracy strategies}
\label{sup:alg:bcd_versions}
\begin{algorithmic}[1]
\REQUIRE Data $\bm X,\bm y$; penalties $R_j,\tau_j$;
mode $\mathsf{Exact}$, $\mathsf{Joint}$ or $\mathsf{Early}$
\STATE Initialize $\bm\beta=\bm0$ and residual $\bm r=\bm y$
\REPEAT
 \FOR{$j=1,\ldots,G$}
  \STATE Form the partial residual, $\bm M_j$ and $\bm z_j$
  \STATE Compute the block update by high-accuracy scalar search
  ($\mathsf{Exact}$) or Algorithm~\ref{sup:alg:normeq_block}
  ($\mathsf{Joint}$ or $\mathsf{Early}$)
  \STATE Store its local KKT bound $\varepsilon_j$ and block increment
  $\bm\Delta_j$; update $\bm r$ and $\bm\beta_j$
 \ENDFOR
 \IF{mode is $\mathsf{Early}$}
   \STATE Compute $\delta_{\rm sw}$ from \eqref{sup:eq:full_sweep_bound}
   \STATE Stop if \eqref{sup:eq:full_early_stop} holds
 \ELSE
   \STATE Evaluate the designated high-accuracy diagnostic
   (the global KKT residual or PG mapping)
   \STATE Stop if its target and any required KKT checks are met
 \ENDIF
\UNTIL{the stopping condition or iteration cap is reached}
\STATE Return the iterate, its residual bound and whether the requested target was met
\end{algorithmic}
\end{algorithm}

In Section~\ref{main-subsec:numerical_sgl}, all six methods target
$\mathcal R_{\rm KKT}\le2\times10^{-5}$.
In Section~\ref{main-subsec:numerical_fsgl}, the three methods target
the common proximal-gradient mapping tolerance $10^{-4}$, with a cap
of 60 BCD sweeps; NormEq--Joint--Dual also checks the nonsmooth KKT
residual at the returned point. The separate stopping experiment in
Section~\ref{main-subsec:sweep_drift} uses NormEq--Joint--CD and records
the first sweep for which $\delta_{\rm sw}\le\xi\lambda_{2,n}$,
for $\xi=0.10,0.05,0.001$. The additional SSLS comparison below
also reports a NormEq--Early implementation at $\xi=0.10$.
Local block tolerances are distinct from the global stopping
tolerances. Warm starts are used across trial scalar values and
successive BCD sweeps.

\subsection{Additional SSLS comparison and group-size scaling}
\label{supsec:ssls_comparison}

We report an additional sparse-group experiment comparing the
configuration-based SSLS method with NormEq. This comparison is
separate from the direct PG/FISTA experiments in the main text.
For the fixed-dimension comparison, we use
$n=200$, $G=20$ and $m=5$ ($p=100$), with independent groups and
within-group correlation $0.75^{|k-\ell|}$. The columns of $\bm X$
are centred and scaled to length $\sqrt n$, and the response is
centred. Four groups are active, including dense and partially sparse
coefficient patterns. We generate
$\bm y=\bm X\bm\beta^*+\bm\varepsilon$ with
$\bm\varepsilon\sim N(\bm0,\bm I_n)$, and use
$\lambda_1=0.04$ and $\lambda_2=0.035$ without tuning for support
recovery. Each method is evaluated on the same 100 paired data sets;
prediction MSE uses 300 independent test observations per replication.
The high-accuracy methods NormEq--Exact, NormEq--Joint and
SSLS--Newton are required to attain the common global KKT target
$2\times10^{-5}$. NormEq--Early is reported at its
regularization-scaled stopping point and is not required to meet
that high-accuracy target.

\begin{table}[htbp]
\centering
\caption{Sparse-group lasso comparison with SSLS at $m=5$
(100 paired replications). Entries are means (standard deviations).
Objective gaps are measured against NormEq--Exact on the same data.
SSLS configuration and root-evaluation counts are accumulated over
the complete fit.}
\label{sup:tab:ssls-baseline}
\small
\setlength{\tabcolsep}{3pt}
\renewcommand{\arraystretch}{1.06}
\textit{Panel A: optimization accuracy and computation}\par\smallskip
\resizebox{\textwidth}{!}{%
\begin{tabular}{lccccc}
\toprule
Method & Time (s) & KKT residual & Objective gap
& Configurations & Root evaluations \\
\midrule
NormEq--Exact
& 1.7215 (0.4126)
& $1.10\times10^{-5}$ ($4.47\times10^{-6}$)
& 0 & -- & -- \\
NormEq--Joint
& 0.1463 (0.0474)
& $1.08\times10^{-5}$ ($4.42\times10^{-6}$)
& $\approx0$ & -- & -- \\
NormEq--Early
& 0.0491 (0.0153)
& $3.99\times10^{-3}$ ($1.79\times10^{-3}$)
& $2.27\times10^{-5}$ ($2.05\times10^{-5}$)
& -- & -- \\
SSLS--Newton
& 0.2823 (0.0924)
& $1.10\times10^{-5}$ ($4.47\times10^{-6}$)
& $\approx0$
& 2488.3 (664.1) & 17389.7 (4642.3) \\
\bottomrule
\end{tabular}}
\medskip

\textit{Panel B: estimation, prediction and support recovery}\par\smallskip
\begin{tabular}{lccc}
\toprule
Method & $\ell_2$ error & Prediction MSE & Support F1 \\
\midrule
NormEq--Exact & 1.3063 (0.2183) & 0.3982 (0.1183) & 0.5540 (0.0727) \\
NormEq--Joint & 1.3063 (0.2183) & 0.3982 (0.1183) & 0.5540 (0.0727) \\
NormEq--Early & 1.3076 (0.2184) & 0.3996 (0.1188) & 0.5506 (0.0723) \\
SSLS--Newton & 1.3063 (0.2183) & 0.3982 (0.1183) & 0.5540 (0.0727) \\
\bottomrule
\end{tabular}
\end{table}

The high-accuracy SSLS--Newton and NormEq fits return essentially
the same estimator and objective value. At this group size,
SSLS--Newton is faster than NormEq--Exact but slower than
NormEq--Joint; NormEq--Early is fastest because it uses a looser,
regularization-scaled stopping rule. The fixed penalties yield
moderate support F1, so this experiment compares computational
procedures rather than tuning for model selection. The root-solver
pilot also shows why we retain Newton as the SSLS representative:
bisection took 0.5247 seconds and 77,966.2 root evaluations per fit,
whereas Newton took 0.2823 seconds and 17,389.7 evaluations.
Newton--ABS took 0.3210 seconds and 26,843.2 evaluations; in this
implementation its additional adaptive-start evaluations outweighed
the saved Newton iterations.

We next examine the effect of group size on the configuration-based
search. We fix $n=500$ and $G=5$, vary
$m\in\{4,5,6,7,8,9\}$, and retain the same within-group covariance
model and penalty parameters as in the fixed-dimension experiment.
There are two active groups, one dense and one partially sparse.
Each setting has 100 paired replications. We record the total number
of SSLS sign/support configurations visited over all block updates
and the per-fit running time of SSLS--Newton, NormEq--Joint and
NormEq--Early.

\begin{table}[htbp]
\centering
\caption{Computational scaling with group size (100 replications
per setting). Entries are means (standard deviations). SSLS
configuration counts sum visits over all block updates and may
include repeated visits; they can therefore exceed $3^m$, the
number of possible configurations for one block update.}
\label{sup:tab:ssls-scaling}
\small
\setlength{\tabcolsep}{4pt}
\renewcommand{\arraystretch}{1.06}
\begin{tabular}{ccccc}
\toprule
$m$ & SSLS configurations & SSLS--Newton (s)
& NormEq--Joint (s) & NormEq--Early (s) \\
\midrule
4 & 36.7 (32.0) & 0.0032 (0.0030) & 0.0103 (0.0043) & 0.0038 (0.0018) \\
5 & 56.9 (78.2) & 0.0055 (0.0079) & 0.0143 (0.0043) & 0.0050 (0.0026) \\
6 & 249.3 (367.9) & 0.0266 (0.0438) & 0.0147 (0.0055) & 0.0052 (0.0022) \\
7 & 521.5 (901.8) & 0.0521 (0.0933) & 0.0164 (0.0048) & 0.0056 (0.0020) \\
8 & 2273.8 (3724.5) & 0.2425 (0.4060) & 0.0225 (0.0072) & 0.0076 (0.0027) \\
9 & 5579.9 (9360.7) & 0.5605 (0.9386) & 0.0241 (0.0092) & 0.0084 (0.0033) \\
\bottomrule
\end{tabular}
\end{table}

\begin{figure}[htbp]
\centering
\includegraphics[width=0.92\textwidth]{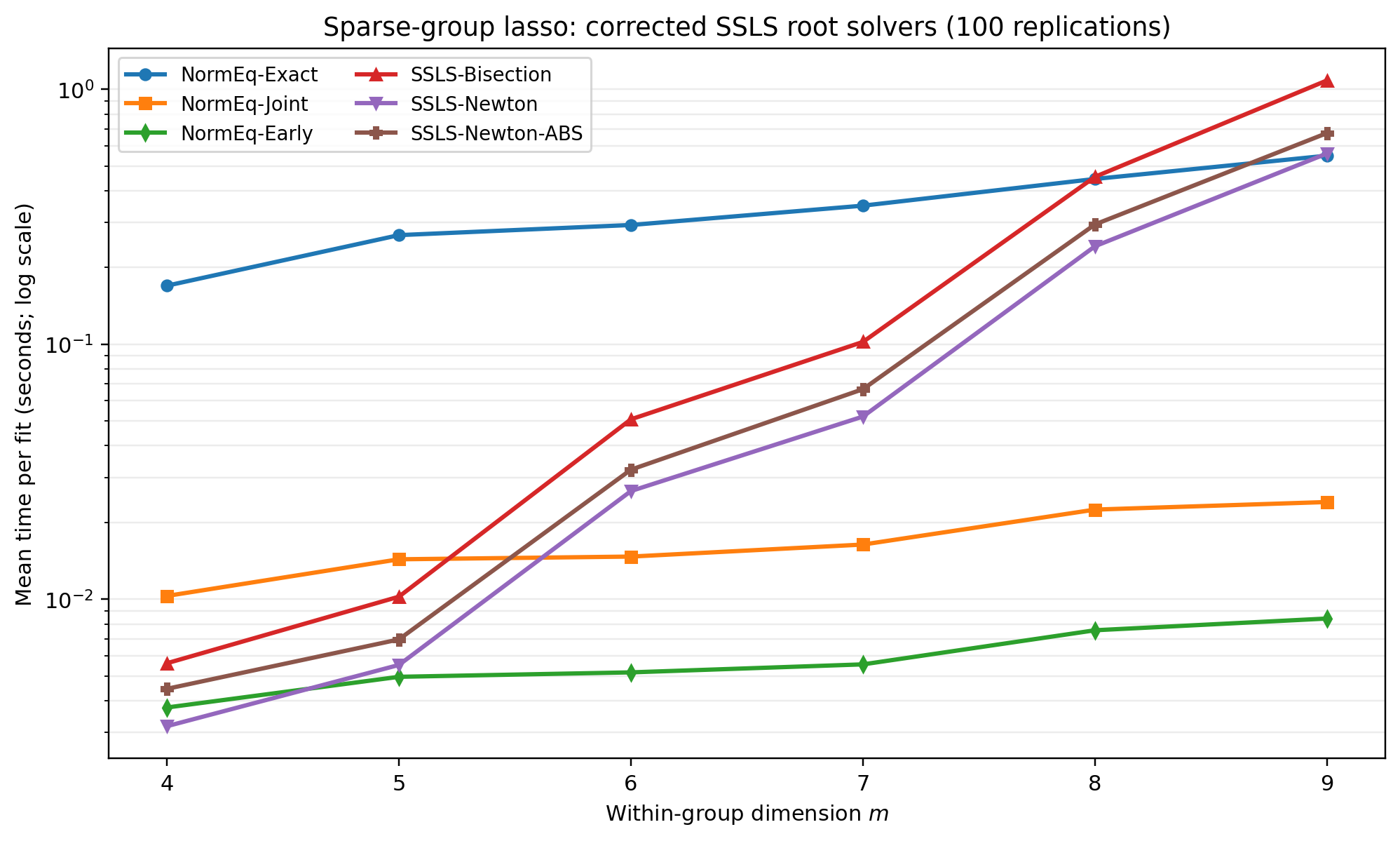}
\caption{Computational scaling with group size. The left panel
shows mean per-fit running times for SSLS--Newton, NormEq--Joint
and NormEq--Early; the right panel shows the mean number of SSLS
configurations visited per fit. Both vertical axes are logarithmic.
Each point is based on 100 paired replications.}
\label{sup:fig:ssls-scaling}
\end{figure}

The mean number of configurations visited over a complete fit
increases from 36.7 at $m=4$ to 5579.9 at $m=9$, while the mean
SSLS--Newton time rises from 0.0032 to 0.5605 seconds. In contrast,
the NormEq--Joint and NormEq--Early times increase from 0.0103 to
0.0241 seconds and from 0.0038 to 0.0084 seconds, respectively.
SSLS--Newton is faster at the smallest group size, but its cost
overtakes NormEq--Joint by $m=6$ and grows much more rapidly thereafter.
The total configuration count is not bounded by $3^m$, because it
accumulates candidate visits across blocks and sweeps; $3^m$ is the
per-block state-space size. These results illustrate the computational
effect of explicit configuration enumeration without implying a
uniform advantage for NormEq at every small group size.

\subsection{Proximal-gradient accuracy and evaluation measures}
\label{supsec:metrics}

For $F_n=f_n+P$ and
\[
t=\{2\lambda_{\max}(\bm X^\top\bm X/n)\}^{-1},
\]
the proximal-gradient mapping norm used in the fused sparse-group
experiment is
\begin{equation}
\label{sup:eq:pg_mapping}
\mathcal R_{\rm PG}(\bm\beta)=
\frac1t\max_j
\left\|\left[\bm\beta-\operatorname{prox}_{tP}
\{\bm\beta-t\nabla f_n(\bm\beta)\}\right]_j\right\|_2.
\end{equation}
Since $P$ is block separable and each block penalty is a Euclidean
group norm plus a within-block penalty, Proposition~\ref{main-prop:isotropic}
gives
\[
\operatorname{prox}_{tP_j}(\bm u)
=\mathbf T_{t\tau_j}\{\operatorname{prox}_{tR_j}(\bm u)\}.
\]
For blocks in the fused sparse-group model, the inner proximal problem
$\operatorname{prox}_{tR_j}$ can be evaluated by a box-constrained
quadratic dual, with bounds $t\lambda_1$ and $t\lambda_f$.
To account for an inexact proximal calculation, let
$\bm u_j=\bm\beta_j-t\nabla_j f_n(\bm\beta)$ and
$H_j(\bm v)=\|\bm v-\bm u_j\|_2^2/2+tR_j(\bm v)$.
Suppose a primal candidate $\widetilde{\bm v}_j$ and a feasible
dual vector $\bm q_j\in t\partial R_j(\bm0)$ yield the
nonnegative primal--dual gap
\[
\mathcal G_j=H_j(\widetilde{\bm v}_j)
-\bigl(\bm u_j^\top\bm q_j-\|\bm q_j\|_2^2/2\bigr).
\]
Strong convexity gives
$\|\widetilde{\bm v}_j-\operatorname{prox}_{tR_j}(\bm u_j)\|_2
\le\sqrt{2\mathcal G_j}$. Since group soft-thresholding is nonexpansive,
with $\widetilde{\bm p}_j=\mathbf T_{t\tau_j}(\widetilde{\bm v}_j)$
we obtain the computable upper bound
\begin{equation}
\label{sup:eq:pg_mapping_gap_bound}
\mathcal R_{\rm PG}(\bm\beta)
\le\frac1t\max_j\left\{
\|\bm\beta_j-\widetilde{\bm p}_j\|_2+\sqrt{2\mathcal G_j}
\right\}.
\end{equation}
This bound is valid when the dual feasibility and gap are evaluated
at the returned proximal candidates. The PG mapping remains distinct
from the block KKT residual evaluated using a valid subgradient
in the norm-equation bounds.

Coefficient error is
$\|\widehat{\bm\beta}-\bm\beta^*\|_2$; prediction MSE is
$\|\bm X_{\rm test}(\widehat{\bm\beta}-\bm\beta^*)\|_2^2/
n_{\rm test}$ using an independent test design.
Support F1 is the harmonic mean of precision and recall on true
nonzero coefficients, using threshold $10^{-5}$ in the supplementary
SSLS comparison of Section~\ref{supsec:ssls_comparison} and $10^{-3}$
in the fused sparse-group experiment of
Section~\ref{main-subsec:numerical_fsgl}. Exact group recovery means
equality of all estimated and true group activity indicators at the
stated threshold. The supplementary SSLS comparison uses 300 test
observations, while the fused sparse-group experiment uses 2500.
The fused sparse-group signal has 40 within-group
edges in the single active $5\times5$ block, and all those true edges
are fused; consequently a true-fusion retention score would not
measure successful separation of distinct levels. Reported times
are per-fit wall-clock times, not the elapsed time for all Monte Carlo
replications run in parallel.

\end{document}